# Analysis of external and internal disorder to understand band-like transport in n-type organic semiconductors

*Marc-Antoine Stoeckel[§], Yoann Olivier[§], Marco Gobbi, Dmytro Dudenko, Vincent Lemaur, Mohamed Zbiri, Anne A. Y. Guilbert, Gabriele D'Avino, Fabiola Liscio, Andrea Migliori, Luca Ortolani, Nicola Demitri, Xin Jin, Young-Gyun Jeong, Andrea Liscio, Marco-Vittorio Nardi, Luca. Pasquali, Luca Razzari, David Beljonne\*, Paolo Samorì\*, Emanuele Orgiu\**

[§] These authors contributed equally to the work

Dr. M.-A. Stoeckel, Dr. M. Gobbi, Prof. P. Samorì, Prof. E. Orgiu
Université de Strasbourg, CNRS, ISIS, 8 allée Gaspard Monge, 67000 Strasbourg, France

Prof. Y. Olivier, Dr. D. Dudenko, Dr. V. Lemaur, Dr. D. Beljonne
Laboratory for Chemistry of Novel Materials, University of Mons, Place du Parc, 20, B-7000 Mons, Belgium

Dr. M. Zbiri
Institut Laue-Langevin, 71 Avenue des Martyrs, 38000 Grenoble, France

Dr. A. A. Y. Guilbert
Centre for Plastic Electronics and Department of Physics, Blackett Laboratory, Imperial College London, London SW7 2AZ, United Kingdom

Dr. G. D'Avino
Institut Néel-CNRS and Université Grenoble Alpes, BP 166, F-38042 Grenoble Cedex 9, France

Dr. F. Liscio, Dr. A. Migliori, Dr. L. Ortolani
CNR - IMM Sezione di Bologna, Via P. Gobetti 101, 40129 Bologna, Italy

Dr. N. Demitri
Elettra - Sincrotrone Trieste, S.S. 14 Km 163.5 in Area Science Park, I-34149 Basovizza, Trieste, Italy

Dr. X. Jin, Dr. Y.-G. Jeon, Prof. L. Razzari, Prof. E. Orgiu
INRS-Centre Énergie Matériaux Télécommunications, 1650 Blv. Lionel-Boulet, J3X 1S2 Varennes, Québec

Dr. A. Liscio
CNR - Institute for Microelectronic and Microsystems (IMM) Section of Roma-CNR, via del fosso del cavaliere 100, 00133 Roma, Italy

Dr. M.-V. Nardi
Istituto dei Materiali per l'Elettronica ed il Magnetismo, IMEM-CNR, Sezione di Trento, Via alla Cascata 56/C, Povo, 38100 Trento, Italy

Prof. L. Pasquali
Istituto Officina dei Materiali, IOM-CNR, s.s. 14, Km. 163.5 in AREA Science Park, 34149 Basovizza, Trieste, Italy

Prof. L. Pasquali

Dipartimento di Ingegneria E. Ferrari, Università di Modena e Reggio Emilia, via Vivarelli 10, 41125 Modena, Italy

Prof. L. Pasquali

Department of Physics, University of Johannesburg, PO Box 524, Auckland Park, 2006, South Africa

E-mail: samori@unistra.fr , david.beljonne@umons.ac.be , emanuele.orgiu@emt.inrs.ca

Abstract: Charge transport in organic semiconductors is notoriously extremely sensitive to the presence of disorder, both internal and external (i.e. related to the interactions with the dielectric layer), especially for n-type materials. Internal dynamic disorder stems from large thermal fluctuations both in intermolecular transfer integrals and (molecular) site energies in weakly interacting van der Waals solids and sources transient localization of the charge carriers. The molecular vibrations that drive transient localization typically operate at low-frequency ($<$ a-few-hundred cm$^{-1}$), which renders it difficult to assess them experimentally. Hitherto, this has prevented the identification of clear molecular design rules to control and reduce dynamic disorder. In addition, the disorder can also be external, being controlled by the gate insulator dielectric properties. Here we report on a comprehensive study of charge transport in two closely related n-type molecular organic semiconductors using a combination of temperature-dependent inelastic neutron scattering and photoelectron spectroscopy corroborated by electrical measurements, theory and simulations. We provide unambiguous evidence that ad hoc molecular design enables to free the electron charge carriers from both internal and external disorder to ultimately reach band-like electron transport.

High charge carrier mobility is a prerequisite to ensure efficient electronic devices such as field-effect transistors (FETs), solar cells, and light-emitting diodes.[1,2] The charge carrier mobility in organic semiconductors results from the interplay of a complex set of physical parameters related to morphology, energetic and structural disorder, and defects, which ultimately are all intimately related the molecular chemical structure.[3] When resorting to organic molecular materials with high degree of crystallinity, the effects of morphology and defects on charge transport are minimized, providing tools to single out the role of disorder. A primary source of disorder is *internal* and stems from large thermal fluctuations of (diagonal) site energies and, mostly, the (off-diagonal) intermolecular interactions mediating transport.[4] The other source of disorder for charge carriers in FETs is *external,* and it arises from the coupling of charge carriers to substrate phonons and randomly oriented dipoles at the semiconductor/gate dielectric interface.[5–7] External disorder was invoked in previous reports as a key limiting factor to charge delocalization.[7] In spite of the above-mentioned internal and external sources of disorder owing to the non-covalent and weak nature of the forces holding together such van der Waals solids, several highly-performing molecular organic single crystals were found to exhibit *band-like transport* near room temperature.[8-11] When properly supported by Hall measurements, the latter transport regime indicates that electron wave functions can be delocalized over several molecular units.[11,12] Although the sole standard electrical FET characterization is not capable of measuring the extent of charge delocalization, the band-like transport is usually invoked when the temperature dependence of the field-effect mobility ($\mu_{FET}$) resembles that observed in inorganic semiconductors such as silicon, i.e. showing an increase in $\mu_{FET}$ upon cooling down from room temperature.

Hence, one can confidently conclude that the combination of internal and external disorder ultimately dictates the extent of charge delocalization in crystalline molecular semiconductors. However, hitherto these two disorder components could never be disentangled. In addition, there is still considerable controversy as to which degree can a carrier wave function extend over neighbouring molecules and especially how this is directly linked to the molecular structure.

Here we use two different perylene diimide (PDI) derivatives as model systems to disentangle the role of the internal vs. external disorder on electron transport in organic crystals. Through a systematic structural, electrical and spectroscopic characterization combining field-effect transistor devices, inelastic neutron scattering, and low-wavenumber spectroscopy measurements together with numerical simulations, we were able to single out the most important vibrational lattice modes (for charge transport) of both PDI single crystals and to correlate them with the molecular chemical structure.

The two PDI derivatives (whose chemical structure is portrayed in **Figure 1a** and **1b**) are functionalized with cyano groups in the bay-region and feature similar environmental stability.[14–17] They only differ by the lateral chain on the imide position, initially designed to enhance their solubility. The first derivative, N-N'-bis(n-octyle)-(1,7&1,6)-dicyanoperylene-3,4:9,10-bis(dicarboximide) known as PDI8-$CN_2$, bears an alkyl chain on each side,[18] while its fluorinated derivative, N-N'-bis(heptafluorobutyl)-(1,7&1,6)-dicyanoperylene-3,4:9,10-bis(dicarboximide) also called as PDIF-$CN_2$, exposes a fluorinated one.[19,20]

Solution-processed single crystals were grown by means of solvent induced precipitation (SIP) (see SI) in order to obtain thin crystals with a reduced number of step edges, which have been found to be detrimental for charge transport.[21] In addition, such a crystal growth method makes it unfavourable for solvent molecules to get incorporated in the final crystal. Once formed, the crystals were then drop-casted on an octadecyl trichlorosilane (OTS)-treated $SiO_2$ substrate. The OTS functionalization acts as a molecular spacer that separates the dielectric surface from the PDI molecules sitting atop. More specifically, such dielectric treatment allows to decrease the coupling between the conjugated electron-transporting part of the molecules and the dielectric[7] and to prevent the presence of silanol groups at the interface of the surface that act as traps for electrons.[22] Two electrodes were evaporated through a shadow mask (Figure S5) to form the final bottom-gate top-contact field-effect transistors (**Figure 1c, d, g**). With this approach, we fabricated top-contact bottom-gate single-crystal FET whose contacts were thermally evaporated through a shadow masks. The latter fabrication step avoids contamination resulting from chemical residues of photoresist and its developer, which may occur when using a standard lithographic process (Figure S6). Furthermore, the dielectric functionalization (operated by functionalizing $SiO_2$ with OTS) allowing to disentangle external vs. internal disorder effects on the electrical transport characteristics can only be employed in a bottom-gate architecture. In addition, the lateral chains acting as a spacer make such contributions negligible for PDI derivatives while this has been attributed as the possible culprit for band-like transport not being observed in pentacene and sexithiophene single crystal FETs.[11] As confirmed by structural characterization, the conjugated core-dielectric distance is nearly identical in both PDIs, which also allows to consider nearly identical the amount of external disorder the two types of molecules are subjected to. Furthermore, $PDI8-CN_2$, and $PDIF-CN_2$ crystals were drop-casted on an OTS layer, which further ensures a minimization of the substrate-induced external disorder.

Both compounds, when integrated into FET structures, present ideal n-type characteristics with reliability factors close to 100% for both PDI compounds, while keeping low maximum power densities and maximum current densities below 4 $W.cm^{-2}$ and 26 $A.cm^{-2}$, respectively (Table S1),[23] with mobility values for $PDIF-CN_2$ being larger than for $PDI8-CN_2$ by about one order of magnitude over the temperature range investigated (**Figure 1f, i**). The trend of the temperature-dependent mobility (Figure 1f) suggests band-like transport in $PDIF-CN_2$, as previously confirmed by Hall measurements.[7] This behaviour could be generally observed in high-purity crystals and it is generally accompanied by an experimental trend of mobility, i.e. $\frac{\partial \mu}{\partial T} < 0$, when charge carrier mobility is not drain-bias dependent.[24,25] Below a given temperature, typically ranging between 180 and 220 K, the transport follows again a thermally-activated mechanism. This is expected since, at such low temperatures, charge transport properties deviate from the ideal behaviour predicted computationally in[26] for defect-free single crystals and instead become dominated by scattering with, and trapping to, defects, as e.g. line (dislocations) or surface (stacking faults) defects. The study of the detailed nature of these defects and how these affect the charge-carrier mobility is beyond the scope of this work. The value of room-temperature mobility measured on a representative statistics of PDIF-CN$_2$ single crystals, approaching 1 cm$^2$V$^{-1}$s$^{-1}$, is very close to previously reported mobility values (~2 cm$^2$V$^{-1}$s$^{-1}$) measured on an insulator with a dielectric permittivity (PMMA, $\varepsilon_r$ ~4) similar to that of our dielectric (SiO$_2$, $\varepsilon_r = 3.9$).[11] However, in few instances, room temperature mobilities 2 cm$^2$V$^{-1}$s$^{-1}$ were also achieved in a previous report on the very same crystals.[27] The apparent mobility mismatch may arise from the high value of the dielectric permittivity employed in our study.[28] Regarding PDI8-CN$_2$ (**Figure 1i**), the transport seems to follow a purely thermally-activated behaviour over the whole measured temperature range (80 K – 300 K). The same electrical behaviour for both derivatives was observed also when a thin insulating layer of BCB, replacing the chemisorbed OTS SAM, was spin-cast on SiO$_2$ (Figure S15 and S16). Since the transport measurements on both PDI derivatives were carried out in three-terminal devices, the contact resistance was determined at each of temperature value at which the mobility was measured (through a classical transmission line method) and the curves corrected accordingly (Figure S12). It is noteworthy that no hysteresis in mobility was observed upon heating and cooling the devices during the measurement (Figure S14). The contact resistance, extracted by following the transmission-line method on single-channel devices integrating crystals of comparable thicknesses (Figure S11) was found to be larger in PDI8-CN$_2$ than in PDIF-CN$_2$ devices. This finding is fully consistent with the characterization performed by means of electron energy loss spectroscopy and ultraviolet photoelectron spectroscopy (**Figure 2**) in single crystals, being therefore fully comparable with the device case, which uses single crystals as the active layer. In particular, EELS measurements of the optical band gap (E$_{opt}$) of the crystals of both compounds provided nearly identical values for PDI8-CN$_2$ and PDIF-CN$_2$. While E$_{opt}$ and presumably the electronic band gap (HOMO-LUMO) of PDI8-CN$_2$ and PDIF-CN$_2$ are nearly equivalent, their respective ionization energies were found to be offset by as much as 0.70 eV (amounting to 7.9 eV and 7.4 eV, for PDIF-CN$_2$ and PDI8-CN$_2$, respectively). Considering the measured energy difference between E$_F$ and HOMO level (Figure 2c), it appears clear that electron injection is certainly more favorable in PDIF-CN$_2$ thanks to the reduction of the energy barrier between Au work function and its LUMO. Structural analysis on the crystals confirmed the bulk structure, slipped-stacked for PDI8-CN$_2$ and brick-wall for PDIF-CN$_2$ (**Figure 3 a-f**), with only one molecule per primitive cell, which is uncommon for small molecule semiconducting materials.[1,13,29]

To rule out any possible phase transition upon temperature, we performed temperature-dependent structural characterizations (**Figure 3 g, h** and Section 4, Supporting Information) and temperature-dependent solid-state F-NMR (Figure S23 and S24) on the fluorine atoms. No significant structural changes (phase transitions) were observed with both techniques upon varying the temperature of the measurement. Thus, we are confident that the observed charge transport behaviours are not driven by any structural change, as it is the case, for example, with tetracene.[30,31]

In order to analyse internal *static* disorder generated by the presence of trap states in the bulk of the crystals, the trap density was extracted in both compounds in the framework of the space-charge-limited current analysis (Figure S17).[32,33] While the absolute value of trap density when extracted through SCLC can sometimes be underestimated, we used this method to compare the ratio of trap density between PDIF-CN$_2$ and PDI8-CN$_2$ single crystals. PDI8-CN$_2$ crystals were found to exhibit a trap bulk density (3 x 10$^{12}$ cm$^{-3}$) comparable to that of PDIF-CN$_2$ crystals (1 x 10$^{13}$ cm$^{-3}$). Our crystals exhibit bulk trap densities which are lower or comparable to those of previously reported vapour-grown single crystals of rubrene (N$_t$ ~ 10$^{15}$ cm$^{-3}$),[33] pentacene (N$_t$ ~ 10$^{11}$ cm$^{-3}$),[34] tetracene (N$_t$ ~ 10$^{13}$ cm$^{-3}$),[31] and solution-grown hydroxycyanobenzene ($N_t \sim 10^{13}$ cm$^{-3}$) single crystals[35] measured through SCLC measurements. However, since in bottom-gate FET devices the charge transport is known to occur within the first nanometers at the interface with the dielectric, a deeper analysis of the electron trapping is required to probe the crystal surface. In particular, step edges at the surface of single crystals were found to trap electrons in both p-type and n-type organic semiconductor crystals.[21] In particular, charge carrier mobilities would decrease with increasing step density. Interestingly, the same study also showed that the thickness of the crystals can influence the type of transport, which can go from thermally activated, in very thick crystals (> 30-µm thick), to band-like transport, in thinner crystals (ca. 5-µm thick). This former experimental observation agrees well with the suggested relationship of proportionality between thickness and density of step edges. More specifically, in the case of PDIF-CN$_2$, the authors demonstrated that charge carrier mobility increases with decreasing the step density. Our AFM measurements carried out on the surface of the very same crystals that were electrically probed in the devices, revealed that the typical thickness of both PDIF-CN$_2$ and PDI8-CN$_2$ crystals falls between 100 and 200 nm. For both derivatives, no step edges were measured on the crystals' surface as revealed by means of KPFM (see section 5, Supporting Information). In this regard, our experimental observation is in line with the discussions by T. He et al. indicating that thin crystals exhibit minimal static disorder thanks to the absence of step edges at which electrons could be trapped. This further experimental evidence allows to state more soundly that the significant differences in transport regimes observed in our molecular systems do not stem from a difference in bulk or surface trap density but should be sought after in the intrinsic (internal) *dynamic* disorder behaviour. Provided that the relevant time scale for charge motion is substantially shorter than the intermolecular vibrational time, charge carriers do transiently (de)localize in space. Hence, the frequency of the intermolecular vibrations is a crucial parameter. In particular, low-frequency (< 200 cm$^{-1}$) large-amplitude vibrations or vibrations displaying strong local and non-local electron-phonon coupling constants are those mostly hampering efficient charge transport. [36]

In order to investigate the charge localization degree of both PDI derivatives, we have performed atomistic calculations of the time-dependent electronic structure of the two molecular crystals by means of a combined classical/quantum modelling approach.[37] Computational details are provided in Supporting Information. In a nutshell, we first ran molecular dynamics (MD) simulations to sample the thermal lattice motion and to compute, as a function of time, the microscopic parameters governing the electron transport in both PDI derivatives. This includes intermolecular charge transfer integrals and molecular site energies, the latter explicitly accounting for supramolecular electrostatic interactions that largely contribute to energetic disorder.[38] We then fed this atomistic information into a tight-binding model for electron states in the two-dimensional high-mobility planes, in order to obtain the localization length (thermally-averaged inverse participation ratio) of electron carriers as a function of time.

The results of the hybrid classical/quantum calculations shown on Figure 3a point to a transient (de)localization of the electron wave function that breathes around average values of ~11 molecules in PDIF-CN$_2$ compared to only ~4 molecules in PDI8-CN$_2$. The large averaged spatial extent of the electron carrier in PDIF-CN$_2$ as opposed to that of PDI8-CN$_2$ suggests a remarkably higher tendency of the carriers to be delocalized in the fluorinated compound. This is a direct consequence of the crystalline packing of both derivatives, being slipped-stack for PDI8-CN$_2$ but brick-wall for PDIF-CN$_2$, as evidenced by the transfer integral distributions in Figure S28.

In order to gain a deeper understanding on the different charge transport mechanisms observed on PDI derivatives, we measured the low-frequency vibrations of both compounds. We employed *temperature-resolved* inelastic neutron scattering (INS) in combination with modelling and simulations. INS makes it possible to explore quantitatively phonon dynamics over the whole Brillouin zone, without being subjected to any specific constraint of selection rules. Unlike very recent works[39,40] based on low-temperature Stokes INS spectra measurements providing insight into molecular vibrations (internal modes), in the present study we used cold-neutron time-of-flight spectroscopy (see section 8 in the SI), in order to measure in the up-scattering regime the anti-Stokes phonon modes (external modes) with a high-energy resolution and an excellent signal-to-noise ratio. Therefore, the low-frequency vibrations (up to 600 cm$^{-1}$) covering both crystal lattice modes (phonons or external modes) and some of the subsequent low-frequency molecular vibrations (internal modes) can be mapped out properly, allowing to study their temperature-dependence (150 - 300K) with direct implications on charge transport (**Figure 4b** and **4c**). In the case of PDI8-CN$_2$, the vibrational peaks become more resolved upon lowering the temperature. This is expected due to a reduction of the thermally-induced displacements, related to the Debye-Waller factor, leading to less broadened features in inelastic neutron scattering. Interestingly, for PDIF-CN$_2$, upon cooling, some low-energy modes below 250 cm$^{-1}$ become clearly distinguishable, exhibiting a pronounced temperature-dependence, seemingly triggering/inducing other effects beyond (or in addition to) the expected reduction of thermal displacements. Further, a new vibrational feature appears at 545 cm$^{-1}$ in the generalized density of states (GDOS).[41] It must be emphasized here that the INS-based GDOS spectra are dominated by the dynamical degrees of freedom of hydrogen atoms, being the strongest neutron scatterers in PDIF-CN$_2$ and PDI8-CN$_2$. Hydrogens are located on the core of the materials in PDIF-CN$_2$ and both on the core and side chains in PDI8-CN$_2$. Therefore, INS intensities of the two PDIs are, in principle, not directly comparable. To underpin our INS measurements, we calculated the INS spectra by Fourier transform of the velocity autocorrelation functions, dissecting the contribution from individual elements along the MD trajectory. The partial contributions had to be neutron-weighted to obtain the GDOS spectra.[40] MD[42] simulations are found to be in a good agreement with the experimental INS spectra by applying a frequency scaling factor of 0.84 and 0.75 for PDI8-CN$_2$ and PDIF-CN$_2$, respectively

(**Figure 4 d, e**). Furthermore, the partial contributions show that the peak around 200 cm$^{-1}$ stems from the nitrogen contribution (Figure S35), with its intensity being larger in the case of PDI8-CN$_2$, and this is also reflected in the room-temperature THz spectroscopy spectrum (Figure S37). Carbon atoms form the host lattice to which the other atoms are bond. Therefore, their dynamical contribution reflects the main lattice vibration in terms of frequency spread. The calculated partial density of states of carbons highlight a stronger intensity of phonons for PDI8-CN$_2$ as compared to PDIF-CN$_2$, up to 192 cm$^{-1}$ (Figure S36). By combining INS with MD simulations, as shown in Figure 4d and 4e, it was possible on the one hand to validate the force field, and on the other hand to get a deeper insight into the partial lattice dynamical contributions. This enabled us to ultimately compare the INS intensities of the two PDIs.

As our model was able to reproduce quite confidently the whole vibrational modes measured by INS, we next proceeded with the calculation of the electron-phonon coupling spectral density for the transfer integrals along the pi-pi direction (see Figure 5a), as the temperature dependence is expected to be mainly sourced by non-local electron-phonon coupling along the dominant transport pathway. Results obtained for the off-diagonal coupling along the pi-edge (see Figure S31) direction as well as for the site energies (Figure S32) are reported in SI. We note that the intense peak around 400 cm$^{-1}$ for PDI8-CN$_2$ is also observed in the electron-phonon coupling spectrum of the site energies (see Figure S30).

These computed spectra convincingly show that low-energy (< 200 cm$^{-1}$) phonons are much more strongly coupled to the electronic degrees of freedom along the pi-pi direction in PDI8-CN$_2$ compared to PDIF-CN$_2$ (**Figure 5b**), thereby rationalizing the more pronounced trend towards breaking of the translational symmetry and spatial confinement of the electrons in the former molecule. The more pronounced intensity of the phonon spectra for PDI8-CN$_2$ was experimentally proved also by room-temperature THz spectroscopy (Figure S37), which allows an immediate and quantitative comparison between the two PDI derivatives. In PDI8-CN$_2$, such low-energy phonons modulate the wave function overlap as a result of a combination of mostly short-axis translations and rotations[43,44] changing the distance and orientation along the packing direction (see SI for detailed assignment), thereby largely affecting the magnitude of the transfer integrals (in contrast with longitudinal displacements). The large (small) relative thermal molecular motions predicted for PDI8-CN$_2$ (PDIF-CN$_2$), as reported in Table S3, translate into broad (narrow) distributions in transfer integrals (Figure S28). Altogether, similarly to didodecyl-benzothienobenzothiophene derivatives[45] and in contrast to the slipped-stack crystal organization of PDI8-CN$_2$, the brick-wall crystal organization in PDIF-CN$_2$ conveys both large electronic interactions and a smaller sensitivity to thermal energetic disorder that act in concert to delocalize the charge carriers, hence triggering band-like transport.

In summary, our systematic and comparative investigation of electron transport in single crystals based on PDI derivatives with subtle difference in the side chains revealed the existence of markedly different charge transport mechanisms, i.e. band-like vs. thermally activated transport. The careful design of our experimental work allowed to experimentally disentangle the effect of external vs. internal (intrinsic) disorder, and to focus on the relationship between the latter and the observed band-like transport through a number of (temperature-resolved) techniques. In particular, charge transport in PDIF-CN$_2$ and PDI8-CN$_2$ was analyzed through temperature-dependent electrical and structural characterization, corroborated by THz spectroscopy and by temperature-dependent inelastic neutron scattering. Our findings are also backed by a sound and fine atomistic modelling, which revealed a more pronounced degree of wave function delocalization for PDIF-CN$_2$ as a result of lower electron-phonon coupling. The experimentally observed differences in transport mechanism between band-like for PDIF-CN$_2$ and thermally-activated transport for PDI8-CN$_2$ can be ascribed to the different phononic activity at low wavenumber, as evidenced by electron-phonon coupling calculations. These calculations show that intrinsic (internal) dynamic disorder builds up in a qualitatively different way in brickwall and slipped-stacked PDI derivatives with respect to herringbone lattices of elongated molecules (e.g. pentacene, BTBT, etc). While in the latter case long-axis intermolecular displacements (below 50 cm$^{-1}$) provide a largely dominant contribution the total energetic disorder,[40] we have revealed here that the scenario is more complex for PDI, with several modes (up to 200 cm$^{-1}$, including short-axis translation and rotations) coming into play. As such, our work suggests rational molecular design guidelines for high mobility (n-type) molecular semiconductors: a combined effect of translations and rotations along the short molecular axis gives rise to low-wavenumber phonons capable of modulating the electron wave function overlap and to affect the magnitude of the transfer integrals in a much more pronounced manner than molecular longitudinal displacements. This study suggests that a general strategy to hinder low frequency phonons can involve the functionalisation of the lateral chains with fluorocarbon groups whose interactions are stronger than based on simple methylation approaches. Such a stronger interaction between -CF$_3$ vs. -CH$_3$ groups can lead to a selective locking of thermal lattice fluctuations below few hundred cm$^{-1}$. We anticipate that an adequate lateral functionalization with fluorocarbons chains along the short axis of brickwall and slipped stack molecular semiconductors (such as indenofluorene[46] or bithiophene-benzothiazole structures[47]) may result into a stiffening of the translational and rotational phonons, leading to smaller amplitude motion at room temperature and therefore a higher degree of delocalisation of the electron wavefunction. While providing a limited set of design rules to achieve the highest degree of electron delocalization in organic semiconductors remains far from trivial, our experimental results coupled with simulations suggests that n-type semiconductors with low-wavenumber phonons associated to short-axis translation generate large energetic disorder. These considerations will therefore help the design of better next-generation electronics materials.

Although this work puts forward the importance of controlling vibrational dynamics through molecular design for electron charge transport in organic FETs, studying and engineering phonons in organic semiconductors is also key for exciton separation in OPVs and the control of heat transport in organic-based thermoelectrics.

**Experimental Section**

*X-Ray Diffraction traces as a function of temperature*

CCDC 1848556, 1848559, 1848555, 1848558, 1848557 and 1848554 contain the supplementary crystallographic data for compounds PDIF-CN$_2$ at 100 K, 150 K, 175 K, 200 K, 250 K and 300 K. These data can be obtained free of charge from The Cambridge Crystallographic Data Centre via https://www.ccdc.cam.ac.uk/structures.

**Acknowledgements**

M.-A. S. and Y.O. contributed equally to the work. E. O. and L. R. are supported by the Natural Sciences and Engineering Research Council of Canada (NSERC) through individual Discovery Grants. E.O. is also supported by the Fonds de Recherche du Québec – Nature et Technologies (FRQNT). The work in Strasbourg was financially supported by EC through the ERC project through the ERC project SUPRAFUNCTION (GA-257305), the Marie Curie ITN projects BORGES (GA No. 813863) and UHMob (GA- 811284), the Labex projects CSC (ANR-10-LABX-0026 CSC) and NIE (ANR-11-LABX-0058 NIE) within the Investissement d'Avenir program ANR-10-IDEX-0002-02, and the International Center for Frontier Research in Chemistry (icFRC). The Institut Laue-Langevin (ILL) facility, Grenoble, France, is acknowledged for providing beam time on the IN6 spectrometer. A. A.Y.G. acknowledges the Engineering and Physical Sciences Research Council (EPSRC) for the award of an EPSRC Postdoctoral Fellowship (EP/P00928X/1). The authors are particularly grateful to Dr. B. Cortese for her scientific assistance with the Scanning Probe Microscopy experiments, and to Dr. J. Raya and Dr. J. Wolf for recording the NMR data. The work in Mons was supported by the European Commission / Région Wallonne (FEDER – BIORGEL project), the Consortium des Équipements de Calcul Intensif (CÉCI), funded by the Fonds National de la Recherche Scientifique (F.R.S.-FNRS) under Grant No. 2.5020.11 as well as the Tier-1 supercomputer of the Fédération Wallonie-Bruxelles, infrastructure funded by the Walloon Region under Grant Agreement n1117545, and FRS-FNRS. The research in Mons is also funded through the European Union Horizon 2020 research and innovation program under Grant Agreement No. 646176 (EXTMOS project). D.B. is a FNRS Research Director. E.O. conceived the research. E.O., P.S. and D.B. supervised the work. M. –A. S. fabricated the devices and performed the electrical measurements under the guidance of M. G. and E. O.. M. –A. S. carried out the solid-state NMR measurements. Y.O., D. O., V. L. and G. D. performed the simulations. A. A. Y. G. and M. Z. conceived the neutron scattering study, wrote the neutron beamtime proposal, carried out the INS experiments, treated and analyzed the data and assisted with data interpretation. F.L. and N. D. carried out the XRD experiments and analyzed the data. F.L., A.M. and L.O. carried out the cryogenic SAED experiments and analyzed the data. A.L. carried out the SPM measurements and analyzed the data. M. V. N. and L. P. performed the UPS and EELS measurements and analyzed the data. X.-J., Y. –G. J. and L. R. carried out the THz measurements and analyzed the data. E.O., M.-A. S. and P.S. wrote the manuscript and all authors participated in manuscript preparation and editing.

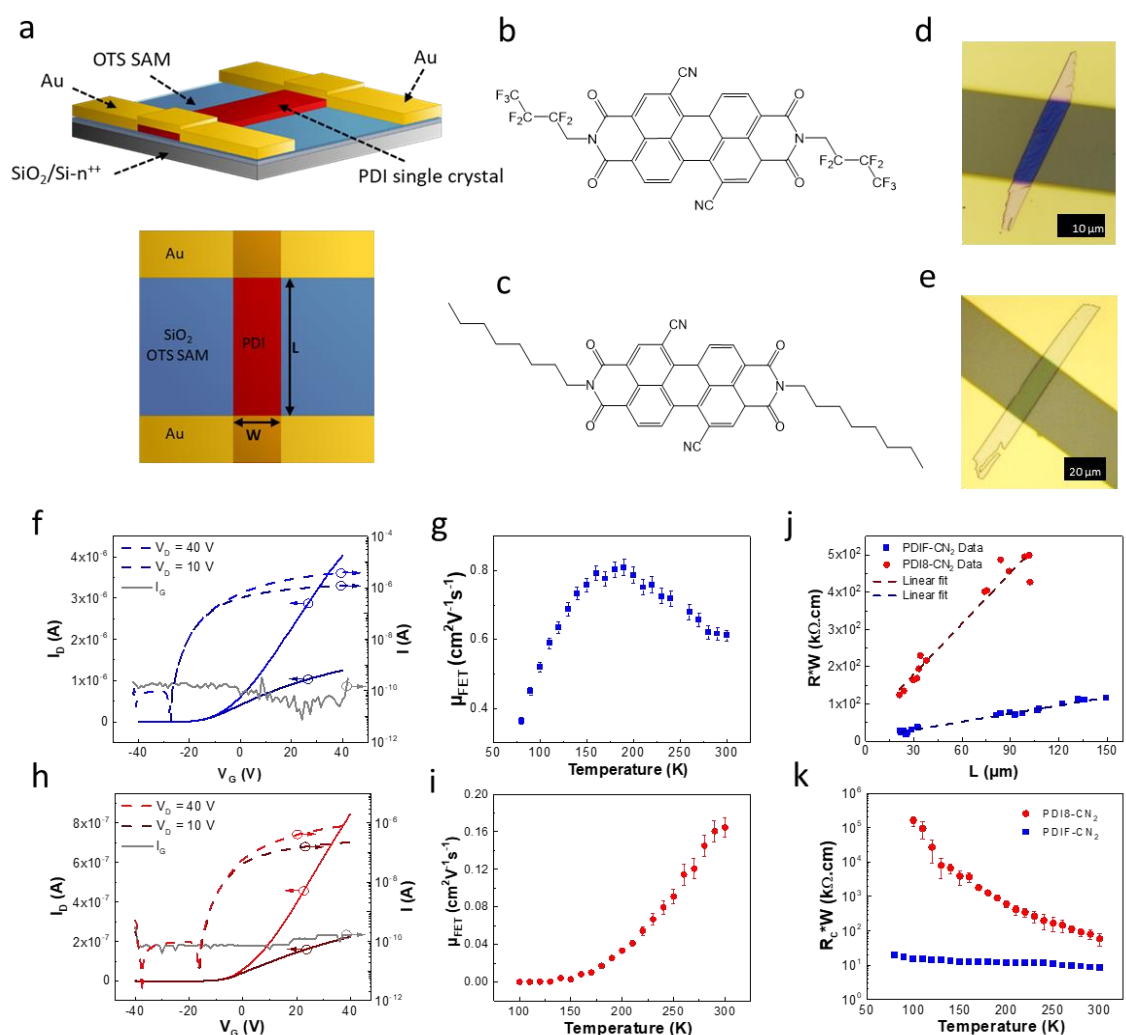

**Figure 1**. **Temperature-dependent electrical characterization of PDI-based devices.**

a) Schematics of a device based on PDI single crystal, 3D and top view. b) chemical structure of PDIF-CN$_2$. c) chemical structure of PDI8-CN$_2$. d) optical microscopy image of a PDIF-CN$_2$ single-crystal FET, f) transfer curve of a single-crystal FET of PDIF-CN$_2$ with V$_D$ = 10 V (in dark blue) and V$_D$ = 40 V (in blue). The gate current is plotted in grey. g) evolution of the linear mobility of a representative PDIF-CN$_2$ single crystal as a function of temperature, exhibiting band-like behavior. e) optical image of a PDI8-CN$_2$ single-crystal FET, h) transfer curve of a PDI8-CN$_2$ single-crystal FET with V$_D$ = 10 V (in dark red) and V$_D$ = 40 V in red. i) evolution of the linear mobility of a representative PDI8-CN$_2$ sample as a function of temperature, revealing thermally-activated transport. j) transmission-line method curves for both PDIF-CN$_2$ and PDI8-CN$_2$ recorded on multiple devices at 300 K, and k) evolution of the contact resistance of both compounds as a function of temperature.

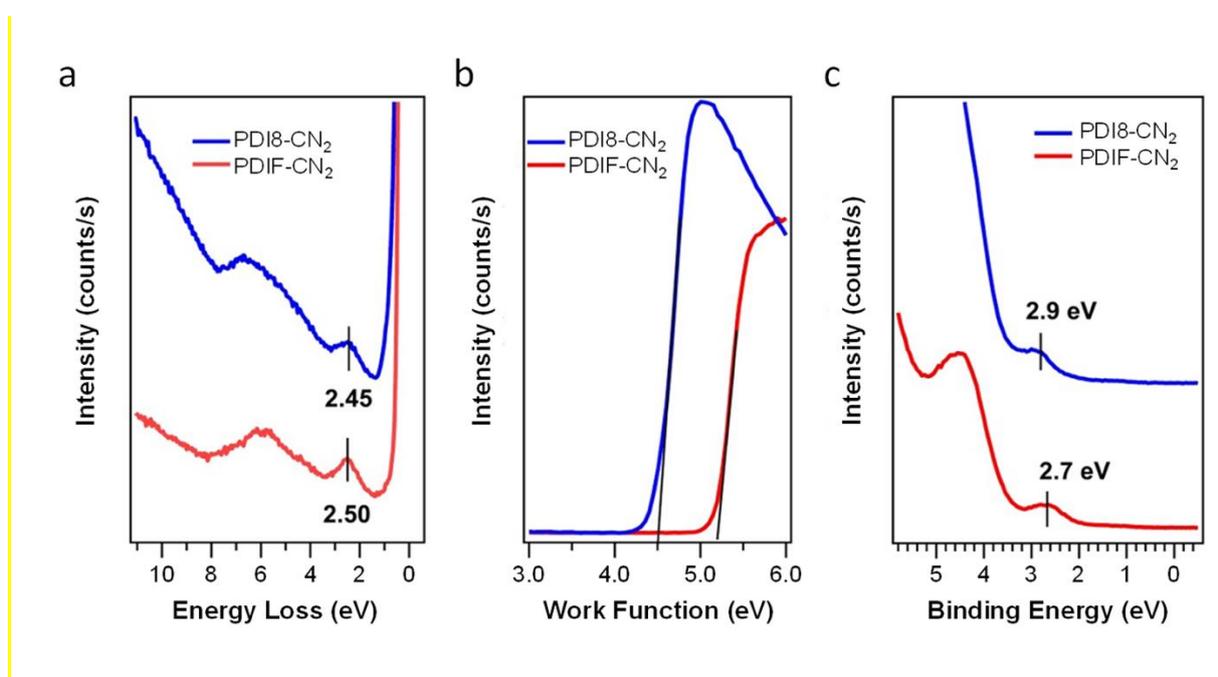

**Figure 2. Energy levels of PDIF-CN$_2$ and PDI8-CN$_2$**
a), EELS spectra of PDIF-CN$_2$ and PDI8-CN$_2$. The marked values in the experimental traces represent the energy difference between the primary beam and the energy of the first loss structure i.e. the optical band gap (E$_{opt}$). E$_{opt}$ corresponds to 2.50 eV for PDIF-CN$_2$ and 2.45 eV for PDI8-CN$_2$. b) and c), work function and E$_F$-HOMO measurements for PDIF-CN$_2$ and PDI8-CN$_2$, respectively.

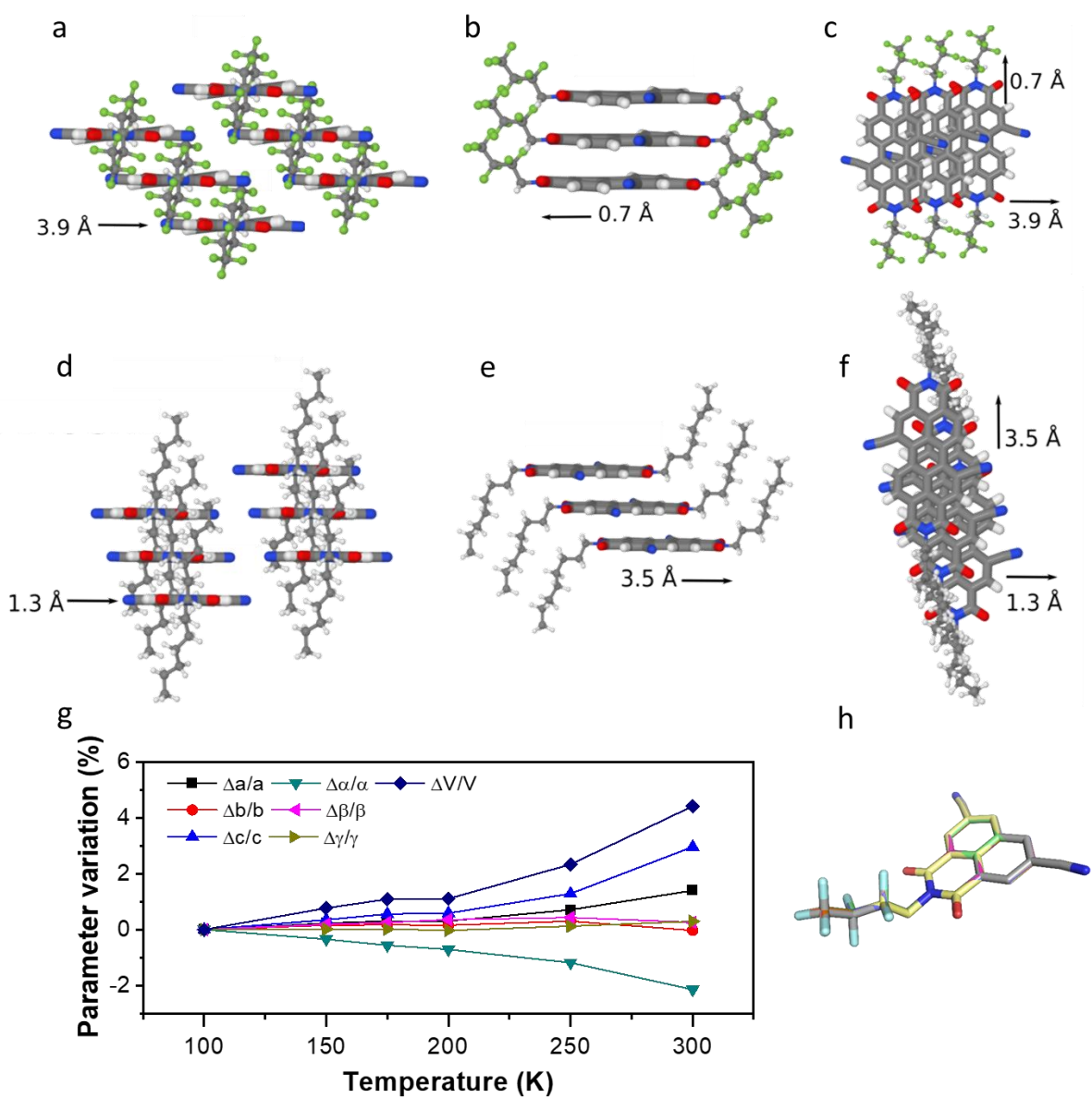

**Figure 3: Structural properties of PDIF-CN$_2$ and PDI8-CN$_2$**
a), b) and c) side view and top view packing of PDIF-CN$_2$. d), e) and f) side view and top view packing of PDI8-CN$_2$. g) variation of the lattice parameter a, b, c, α, β, γ, and volume (V) during the thermal annealing of PDIF-CN$_2$ single crystals where strongest variation was the expansion by 3% of the c-axis at 300 K, which lead the volume to increase by 4.4%, which is not indicative of an actual phase transition. h), Overlap of equivalent PDIF-CN$_2$ conformations found in 100 K – 300 K models (R.M.S.D. between models < 0.05 Å). Hydrogens omitted for clarity.

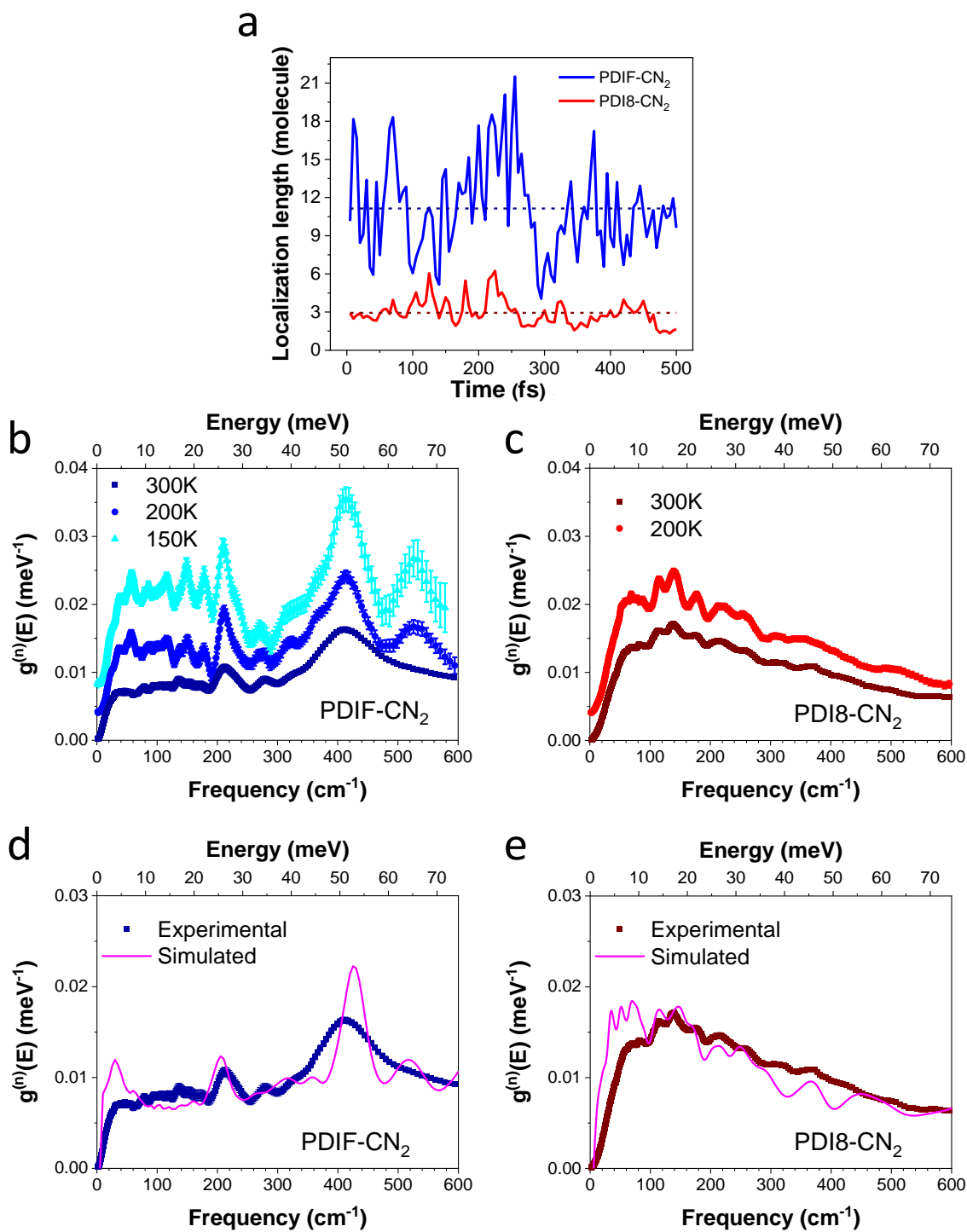

**Figure 4: Experimentally determined and calculated vibrational modes of PDIF-CN$_2$ and PDI8-CN$_2$**

a) Localization length over time: charge carriers are delocalized over ca. 11 PDIF-CN$_2$ molecules (in blue). Conversely, charge carriers are localized only over ca. 4 molecular units in the case of PDI8-CN$_2$ (in red). b) and c) temperature-dependent inelastic neutron scattering spectra in terms of the GDOS of PDIF-CN$_2$ and PDI8-CN$_2$ respectively. (Full GDOS spectra resolved in temperature shown in S33); d) and e) comparative experimental-simulated INS-based GDOS spectra of PDIF-CN$_2$ and PDI8-CN$_2$, respectively (room temperature).

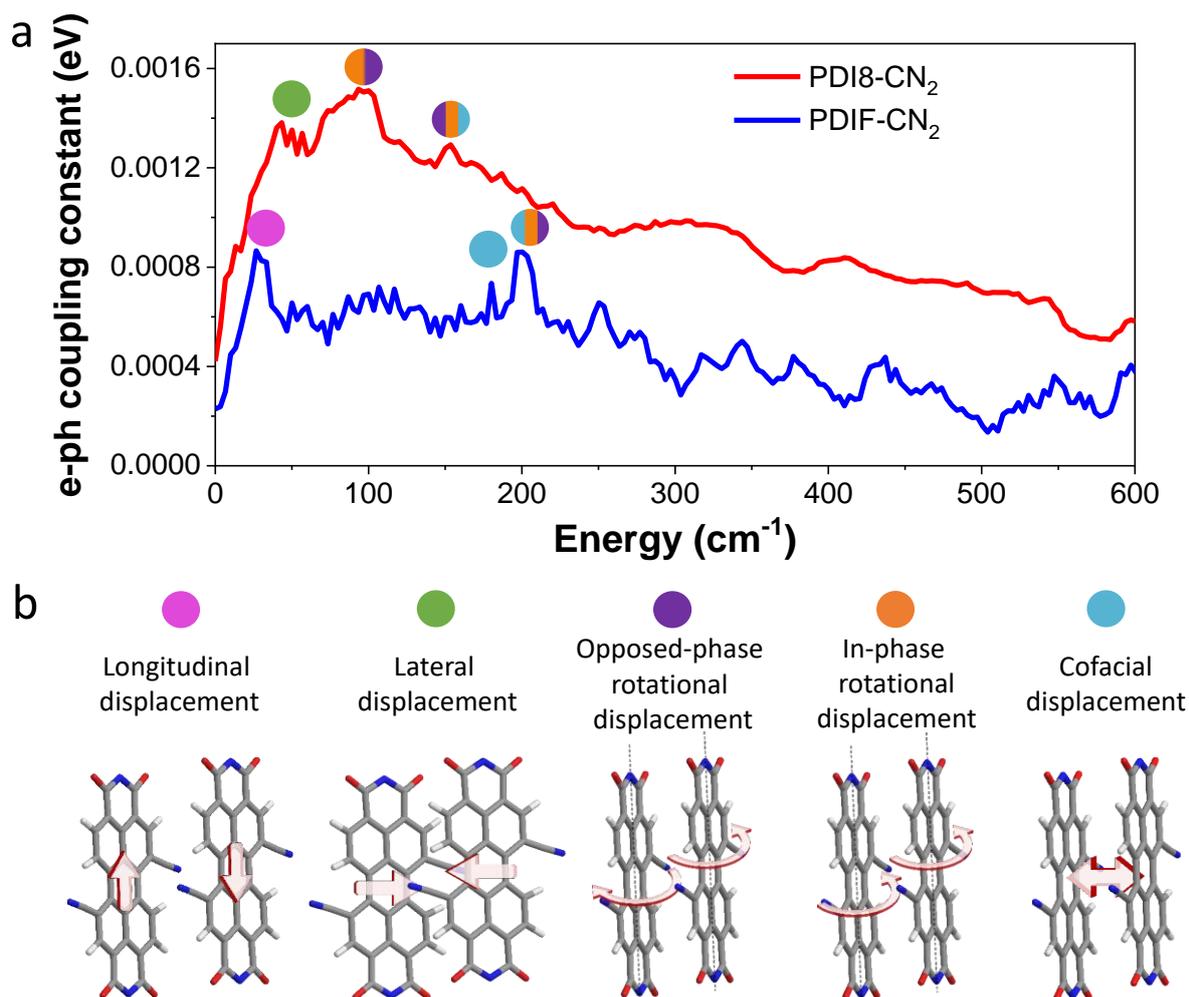

**Figure 5: Electron-phonon coupling analysis and assigned molecular displacement for PDI8-CN$_2$ and PDIF-CN$_2$**

a) Electron-phonon coupling spectrum of the time-dependent transfer integrals along the pi-pi direction in crystals of PDI8-CN$_2$ (in red) and PDIF-CN$_2$ (in blue), respectively (T = T$_{amb}$). b) Different molecular displacements and related color code associated to the spectra in (a).

Molecular vibrations govern the charge transport in organic semiconductors which is limited by different sources of disorder. Understanding and mastering the disorder in these materials can drive the design of better semiconductors featuring band-like transport.

**Keyword** organic semiconductors, phonons, charge transport, field-effect transistors, disorder

Marc-Antoine Stoeckel, Yoann Olivier, Marco Gobbi, Dmytro Dudenko, Vincent Lemaur, Mohamed Zbiri, Anne A. Y. Guilbert, Gabriele D'Avino, Fabiola Liscio, Andrea Migliori, Luca Ortolani, Nicola Demitri, Xin Jin, Young-Gyun Jeong, Andrea Liscio, Marco-Vittorio Nardi, Luca. Pasquali, Luca Razzari, David Beljonne\*, Paolo Samorì\*, Emanuele Orgiu\*

**Analysis of external and internal disorder to understand band-like transport in n-type organic semiconductors**

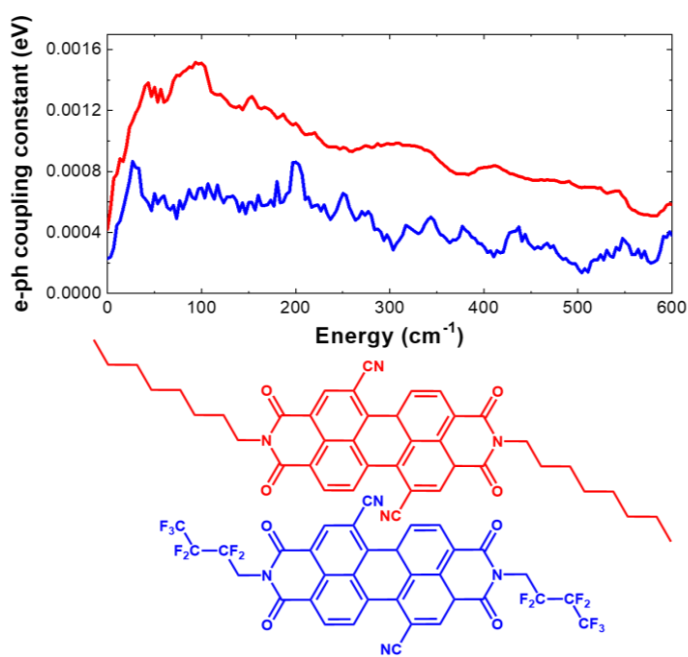

Supporting Information

# Analysis of external and internal disorder to understand band-like transport in n-type organic semiconductors

*Marc-Antoine Stoeckel, Yoann Olivier, Marco Gobbi, Dmytro Dudenko, Vincent Lemaur, Mohamed Zbiri, Anne A. Y. Guilbert, Gabriele D'Avino, Fabiola Liscio, Andrea Migliori, Luca Ortolani, Nicola Demitri, Xin Jin, Young-Gyun Jeong, Andrea Liscio, Marco-Vittorio Nardi, Luca. Pasquali, Luca Razzari, David Beljonne\*, Paolo Samorì\*, Emanuele Orgiu\**

## 1. Optical absorption characterization

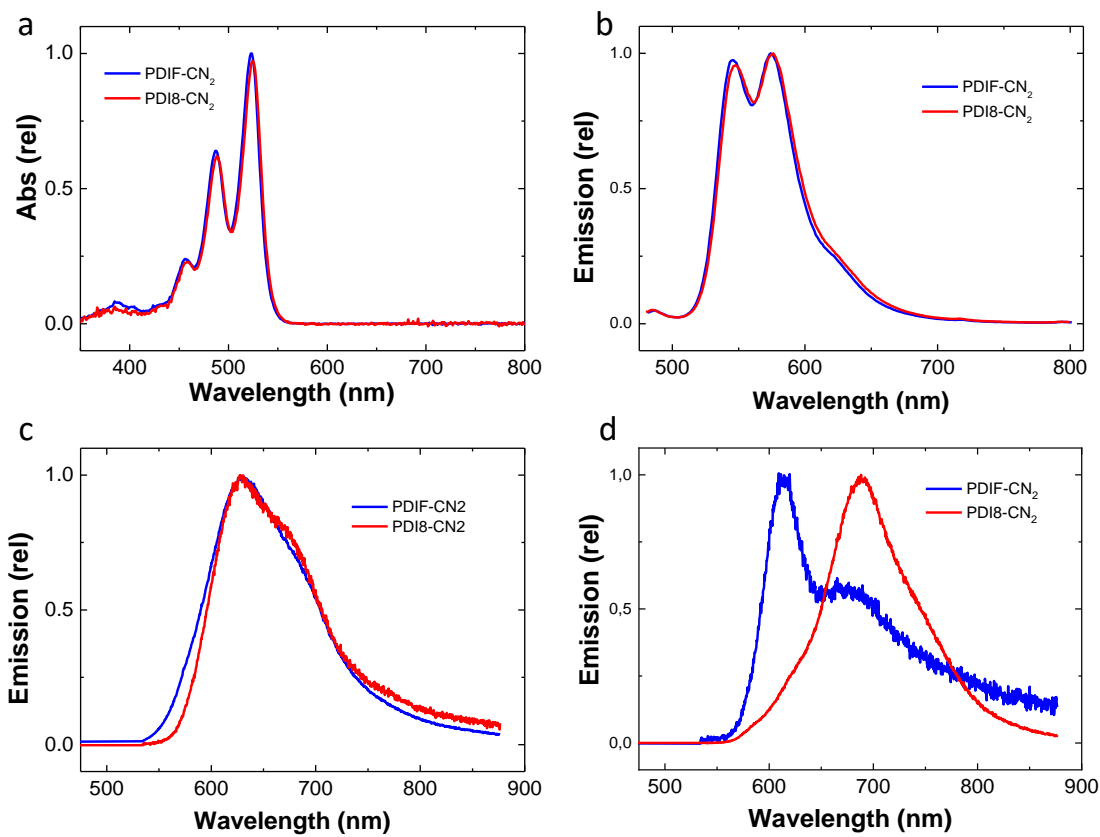

**Figure S1: a**, Normalized UV-vis absorbance spectra of PDIF-CN$_2$ and PDI8-CN$_2$ in CHCl$_3$ (0.002 mg/mL) and **b**, their related emission spectra. **c**, normalized emission spectra of spin-coated PDIF-CN$_2$ and PDI8-CN$_2$ films on SiO$_2$ substrates (2 mg/mL in CHCl$_3$) and **d**, normalized emission spectra of PDIF-CN$_2$ and PDI8-CN$_2$ single crystal.

## 2. Crystal growth and device fabrication

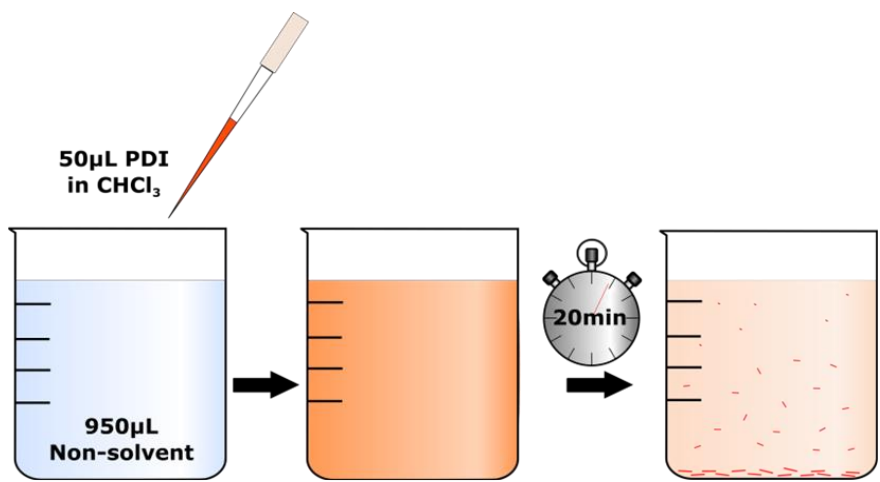

**Figure S2**: Principle of the solvent induced precipitation (SIP) method. 50 μL of PDI (Polyera Corporation) solution at 1 mg/mL in chloroform are poured into 950 μL of a non-solvent for PDI (acetone for PDI8-CN$_2$, methanol for PDIF-CN$_2$). After 20 min, large crystals are formed and precipitate at the bottom of the vial, while smaller crystals are suspended in the middle of the vial. Longer crystals are formed by heating up at 45 °C the mixture of non-solvent and PDI.

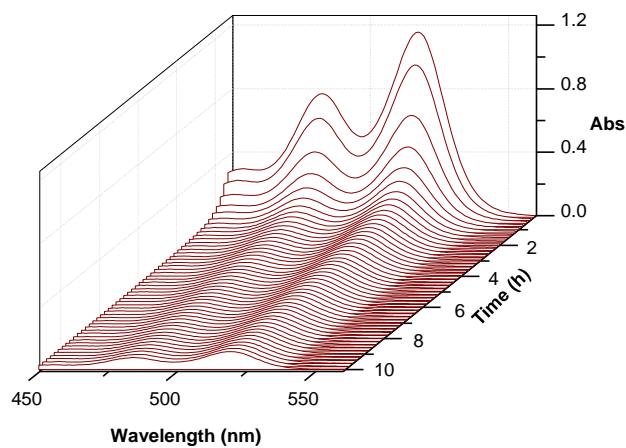

**Figure S3**: **a**, UV-vis spectra recorded for PDIF-CN$_2$ in hot methanol (45 °C) at the beginning of the cooling process, to follow the formation of the crystals that leads to a decrease of the absorbance. Large crystals would form typically after 150 minutes.

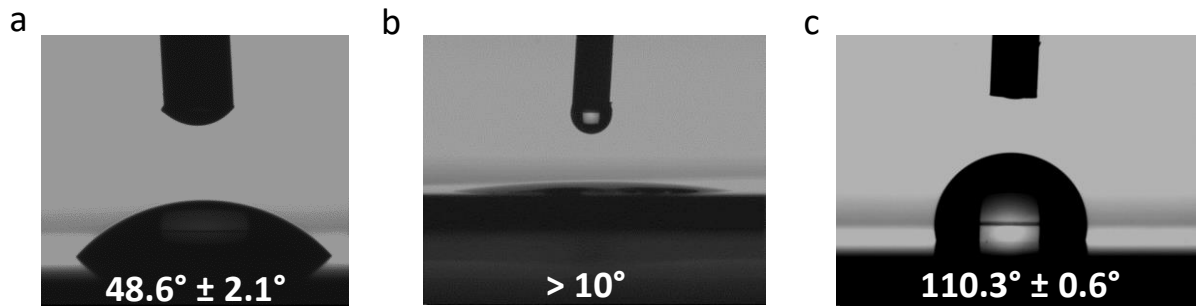

**Figure S4:** Water contact angle characterization of **a**, thermally growth SiO$_2$,[1] **b**, UV/O$_3$ treated SiO$_2$ for 5 min of exposition and 25 min of incubation, leading to an increase of the wettability due to the formation of hydroxyl groups at the surface[2] and **c**, an octadecyltrichlorosilane (OTS) treated SiO$_2$ that decreases the wettability induced by the alkyl chains on the surface.[3]

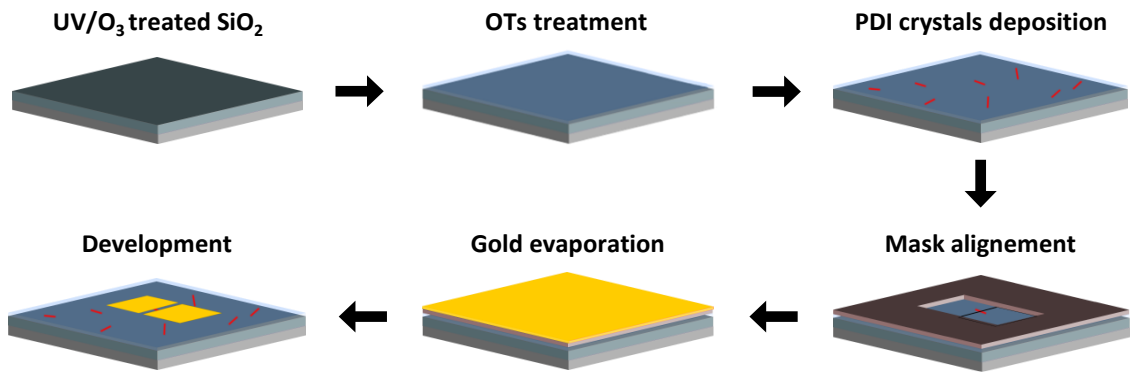

**Figure S5**: Schematics of the fabrications of PDIs single-crystal FET. SiO$_2$/Si-n$^{++}$ substrates 230-nm thick SiO$_2$ (from Fraunhofer Institute for Photonic Microsystems IPMS, Dresden, Germany) were cleaned in an ultrasonic bath of acetone then propan-2-ol. The substrates were then gently dried with a nitrogen flow.

*Deposition of SIP-grown crystals*: 20 μL of the solution containing the PDIs crystals are taken from the vial and dropcasted on the substrate. A shadow mask made with different L (25 μm to 150 μm) was then aligned on a single crystal under an optical microscope. 40 nm of gold were evaporated through this mask, forming two electrodes over the targeted crystal.

*Device fabrication and characterization*

SiO$_2$/Si-n$^{++}$ substrates 230-nm thick SiO$_2$ (purchased from Fraunhofer Institute for Photonic Microsystems IPMS, Dresden, Germany) were first cleaned in ultrasonic bath of acetone followed by a further sonication cycle in propan-2-ol. The substrates were then dried with gentle flow of nitrogen.

*OTS SAM preparation procedure:* To form the OTS SAM layer, the substrates are first exposed to a UV/Ozone atmosphere for 30 min then are immediately placed inside a glovebox filled with N$_2$ atmosphere, and soaked in a solution of 10 mL of anhydrous toluene and 40 µL of octadecyltrichlorosilane (OTS, Sigma Aldrich) in a sealed glass jar. The jar is then heated at 60 °C for 1h and left undisturbed overnight at room temperature.

*BCB dielectric deposition procedure*: To form a BCB layer, the monomer (Cyclotene 3022-46, Dow) was mixed with Mesitylene (Sigma-Aldrich) at a 1:4 vol. ratio, and the final solution was stirred for 30 min. In a glovebox filled with nitrogen, 150 µL of this solution were spin-coated on SiO$_2$ surface (2500 rpm, 4000 rpm/sec, 60 sec). The as-prepared films were then annealed at 110 °C for 10 min to remove residual solvent. This step was followed by a second annealing cycle at 290 °C for 10 min for cross-linking. The BCB film thickness was determined to be 150 nm (measured with an Alpha-Step IQ profilometer form Kla Tencor). BCB has a dielectric constant of 2.65, leading to a final BCB/SiO$_2$ bilayer capacitance of 7.66 nF/cm$^2$.

After that, the substrates are removed from solution, rinsed twice in anhydrous toluene following annealing for 1 h at 60 °C. 50 µL of PDI (Polyera Corporation) solution at 1 mg/mL in chloroform (a good solvent for PDIs) are poured into 950 µL of a non-solvent for PDI (acetone for PDI8-CN$_2$, methanol for PDIF-CN$_2$). After 20 min, large crystals are formed and precipitate at the bottom of the vial, while smallest crystals are suspended in the middle of the vial. 20 µL of this solution is then taken from the bottom of the vial and drop-casted on non-patterned substrates. A homemade shadow mask made with metallic wires of different width (25 μm to 150 μm) is placed on a single crystal, with the help of a microscope. 40 nm of gold are evaporated through this mask, forming two electrodes over the targeted crystal.

*Low-temperature electrical measurements*

Electrical measurements at low temperature were done with a Nitrogen cryostat Oxford Instrument Optistat DN-V from 80 K to 300 K in steps of 10 K, controlled through an ITC503S Cryogenic Temperature Controller and interfaced with a dual-channel Keithley 2636A source-meter. The classical transistor transfer curves were acquired by sweeping the gate voltage ($V_G$) between -40 V to 40 V when applying a drain voltage ($V_D$) of 10 V and 40 V applied with a dual channel Keithley 2636A source-meter synchronized with the cryostat through an in-house Labview script. More than 20 devices per PDI derivative were electrically probed.

*Field-effect mobility measurements*

The electron mobility was calculated in the linear regime according to the following equation:

$$\mu_{Lin} = \left(\frac{\partial I_D}{\partial V_G}\right)_{V_D} \cdot \left(\frac{L}{W}\right) \cdot \left(\frac{1}{C_i}\right) \cdot \left(\frac{1}{V_D}\right)$$

where $V_D$ and $I_D$ are the potential applied and the current measured respectively between the source and the drain electrodes, $V_G$ is the potential difference applied between source and gate electrode, L is the source-drain electrode distance, W is the channel width and $C_i$ the capacitance per unit area of the insulator layer.

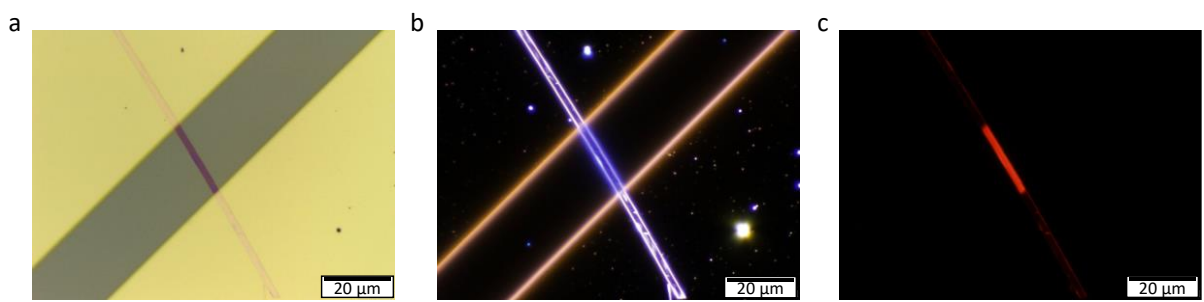

**Figure S6**: Typical PDI based single-crystal FET fabricated. The optical image in **a**, represents the measured single-crystal in bright field, **b** in dark field and, **c**, in excitation mode ($\lambda_{ex}$ = 450 nm).

## 3. Temperature-dependent device characterization

|  | $r_{lin}$ (%) | $P_{max}$ (W.cm$^{-2}$) | $J_{SD}^{max}$ (A.cm$^{-2}$) |
|---|---|---|---|
| *PDIF-CN$_2$* | 92.9 | 3.12 | 25.7 |
| *PDI8-CN$_2$* | 104 | 0.63 | 7.38 |

**Table S1:** Reliability factor in linear regime $r_{lin}$, maximum power density $P_{max}$ and maximum current density $J_{SD}^{max}$ calculated for PDIF-CN$_2$ and PDI8-CN$_2$.[4] All mobilities measurements were carried out according to Choi et al. Critical assessment of charge mobility extraction in FETs. *Nat. Mater.* **17,** 2 (2017).

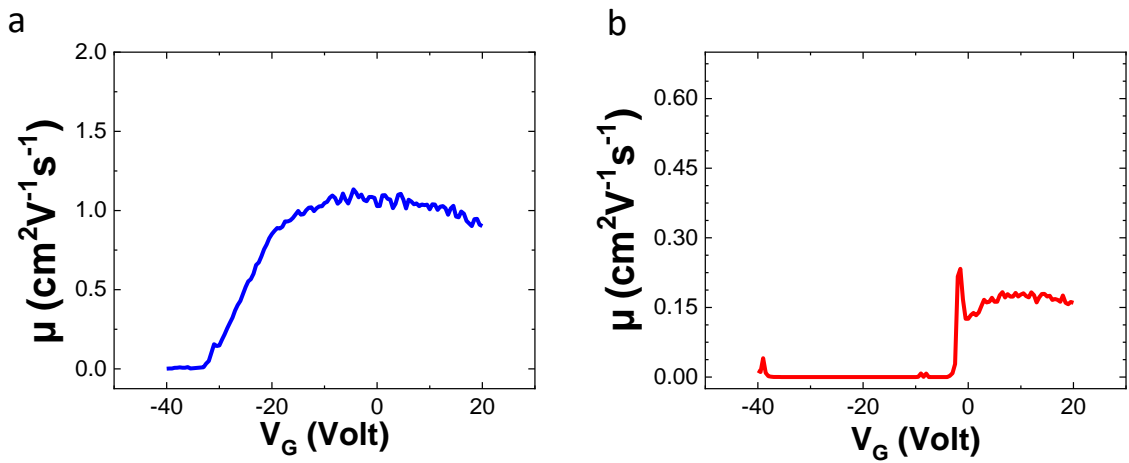

**Figure S7:** Charge carrier mobility vs. gate voltage for **a**, PDIF-CN$_2$ and **b**, PDI8-CN$_2$, recorded at 300 K ($V_D$ = 40 V). In both cases, the charge carrier mobility is rather stable above the threshold (gate) voltage.

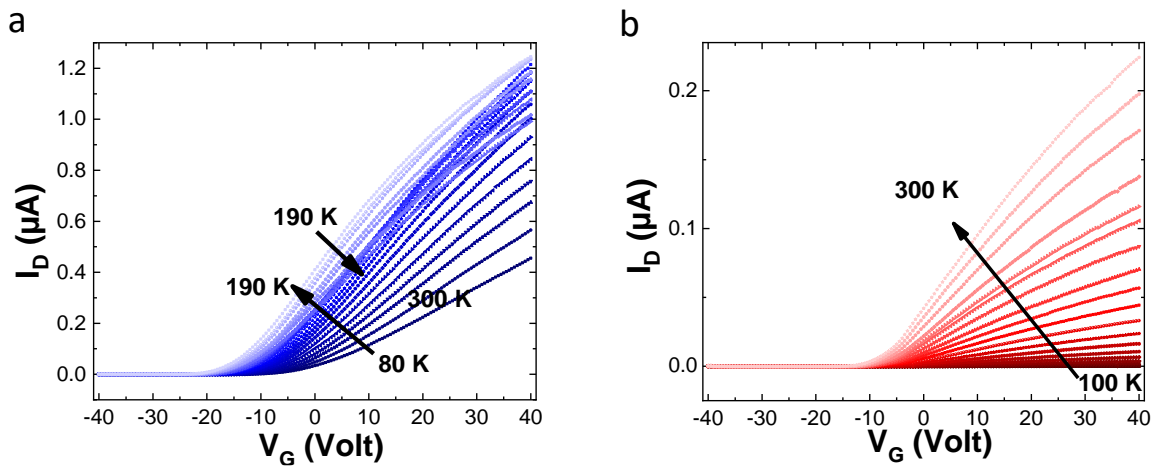

**Figure S8:** Transfer curves at between 80 K and 300 K in **a**, PDIF-CN$_2$ and in **b** for PDI8-CN$_2$.

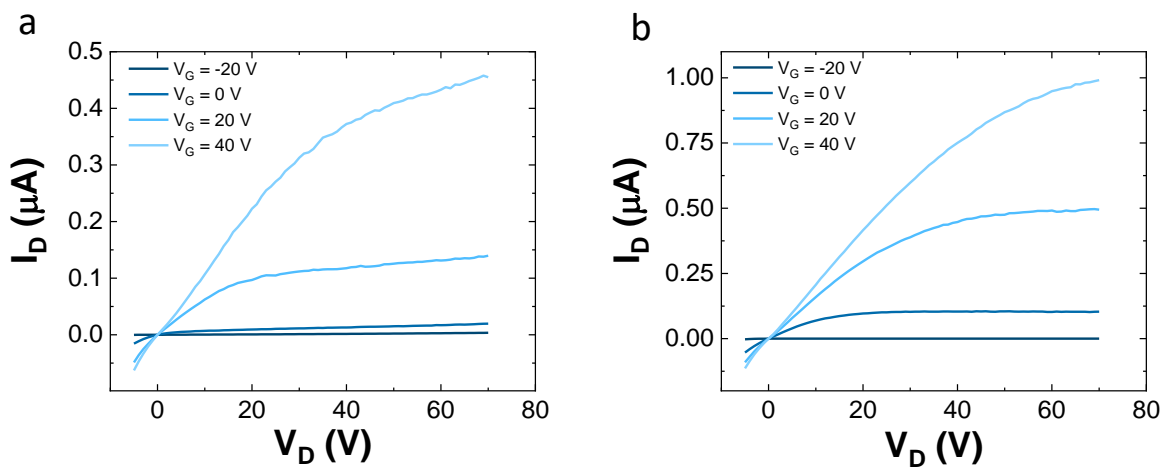

**Figure S9:** Output curves recorded at **a**, 80 K and **b** 300 K for PDIF-CN$_2$.

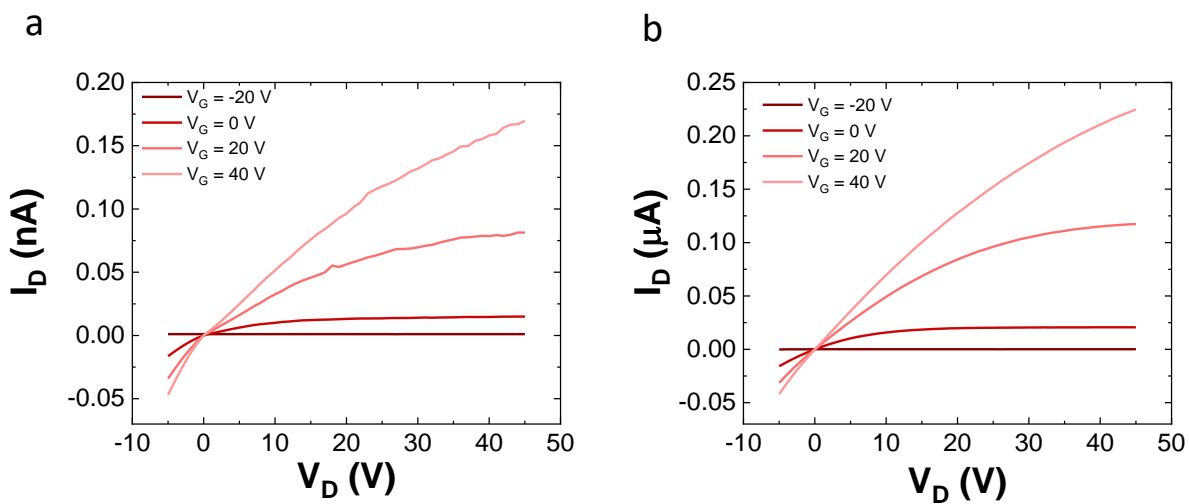

**Figure S10:** Output curves recorded at **a**, 120 K and **b** 300 K for PDI8-CN$_2$.

*Transmission line method*

To account for contact resistance effect in FET devices, we applied the well-known transmission line method (TLM). To apply this method, we measured transistor output characteristics with different channel length where the resistance of the linear regime is extracted. The resistance measured R is the sum of the channel resistance $R_{Ch}$, which is dependent to the channel length, and the metal-semiconductor contact resistance, $R_C$, which is independent to the channel length, can be expressed as:

$$R = R_{Ch} + R_C = \frac{L}{W\mu C(V_G - V_T)} + R_C$$

Where L is the channel length of the device and W its channel width, µ is the charge carrier mobility C is the capacitance per unit area, $V_G$ and $V_T$ are the gate and the threshold voltage respectively. When the measured resistance R is plotted as a function of the channel length, the contact resistance $R_C$ can be extracted at L = 0 µm. The resistance is width corrected as the width of the crystals is slightly fluctuating from one device to another. The contact resistance is then extracted for each temperature, and the measured mobilities for each device separately, are corrected with it. The statistics on mobilities plotted Fig 1 in the main text results from these corrected individually corrected mobilities. No gate voltage dependence of the contact resistance was evidenced in our measurements.

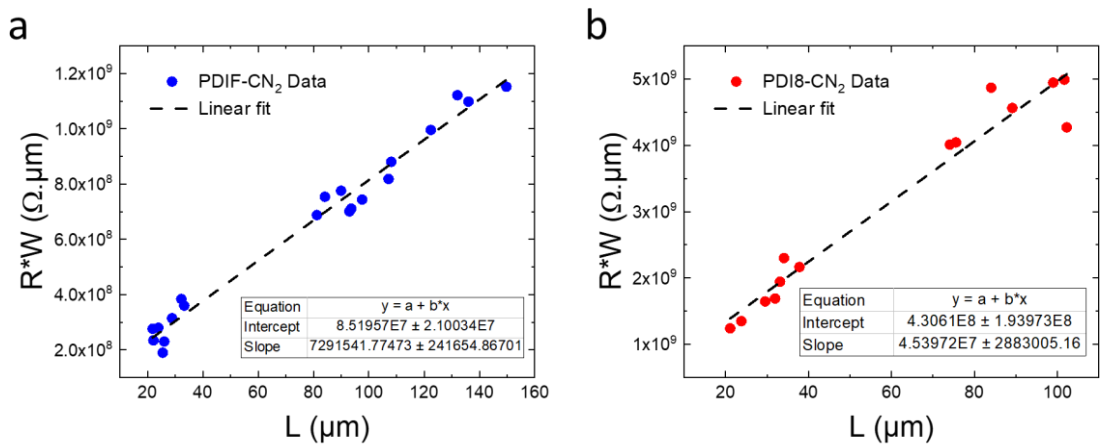

**Figure S11:** *Transmission-line method (TLM).* Channel width-corrected resistance versus channel length of **a**, PDIF-$CN_2$ and **b**, PDI8-$CN_2$ single crystals recorded at T = 300 K, on multiples devices with comparable thicknesses.

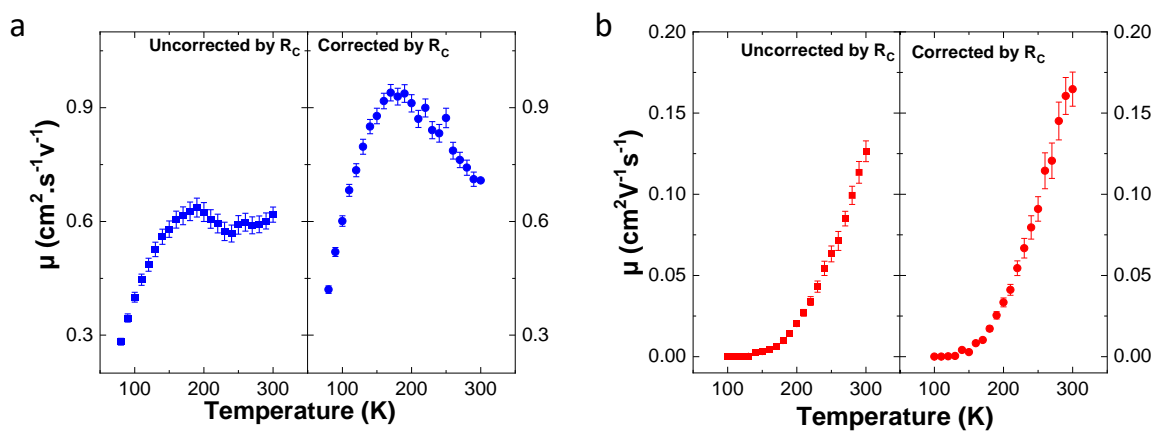

**Figure S12:** *Correction of charge carrier mobility by contact resistance through Transmission Line Method.* **a**, statistics on charge carrier mobility (linear regime, std. err.) vs. temperature for PDIF-$CN_2$ with and without the contact resistance correction and the respective one in **b** for PDI8-$CN_2$. We noticed that the contact resistance was found to be higher for PDI8-$CN_2$ than for PDIF-$CN_2$, which is consistent with the experimental energy level diagrams presented later.

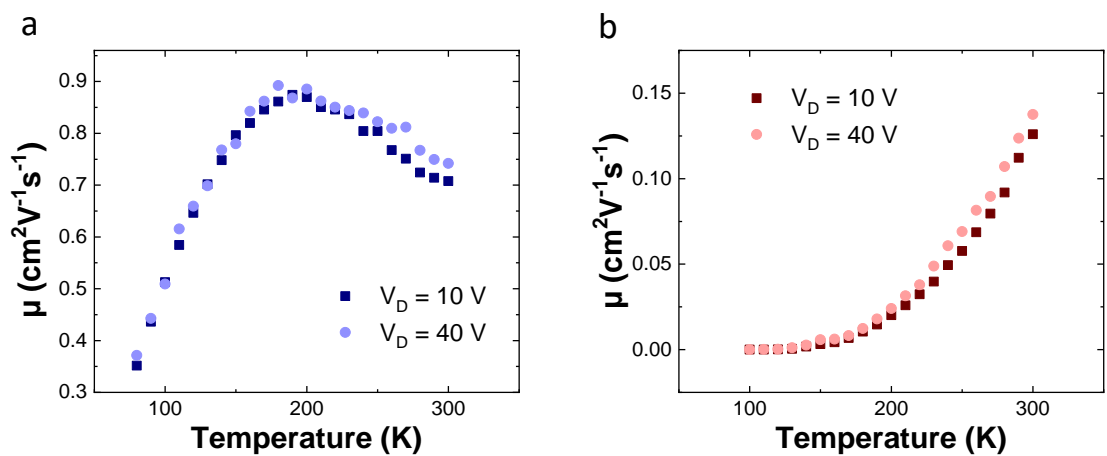

**Figure S13:** Charge carrier mobility at different drain voltages vs. temperature for **a**, PDIF-$CN_2$ and **b**, PDI8-$CN_2$. The charge carrier mobility does not strongly depend on $V_D$ in either compounds.

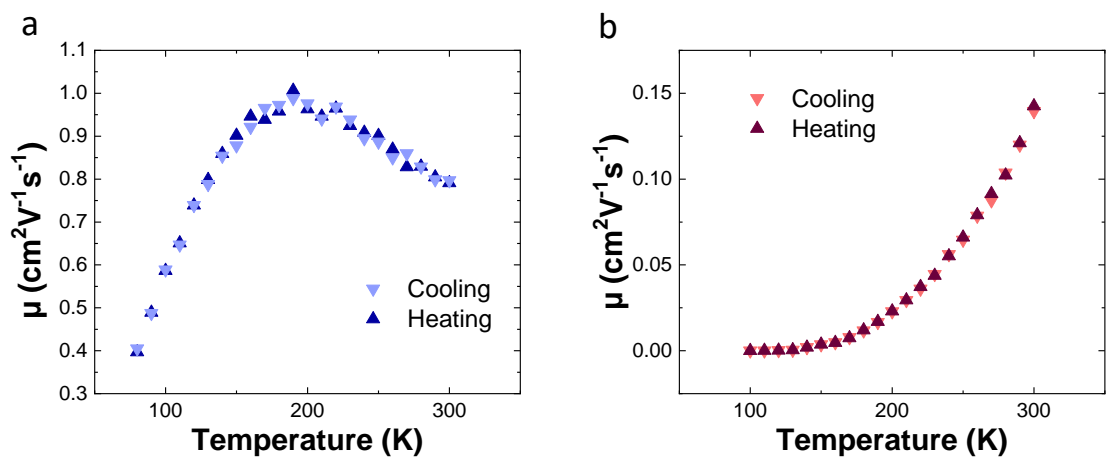

**Figure S14:** Charge carrier mobility vs. temperature for **a**, PDIF-$CN_2$ and **b**, PDI8-$CN_2$ upon cooling then heating, showing no signs of hysteresis.

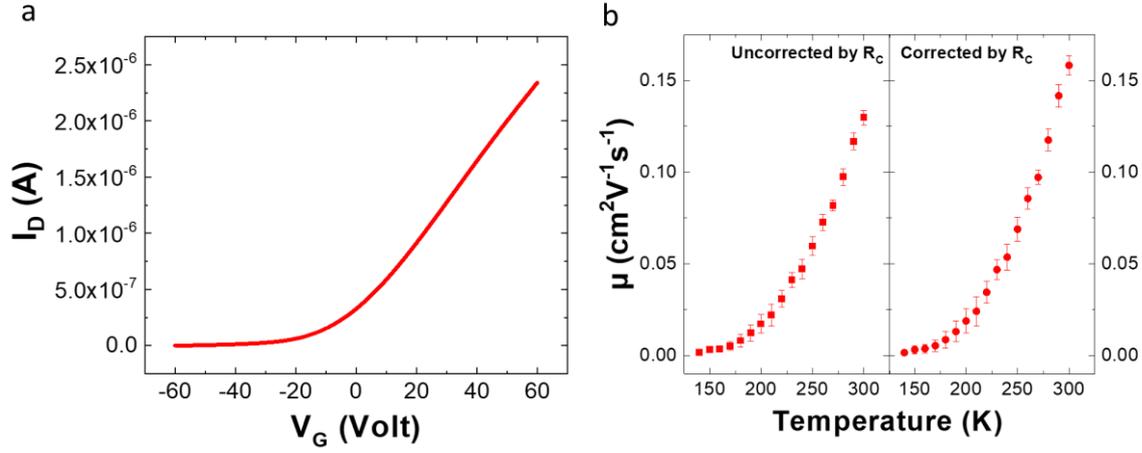

**Figure S15: a**, typical transfer curve ($V_D = 40$ V, T = 300 K) recorded for single-crystal PDI8-CN$_2$ FET on a BCB/SiO$_2$ dielectric. **b**, statistics (std. err. over 3 devices) on charge carrier mobility vs. temperature for PDI8-CN$_2$, with and without contact resistance correction.

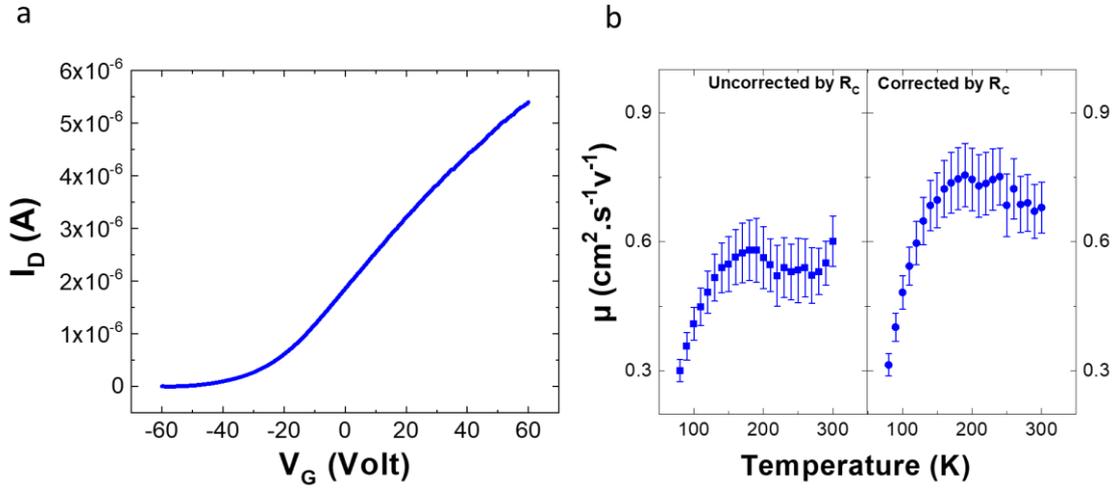

**Figure S16: a**, typical transfer curve ($V_D = 40$ V, T = 300 K) recorded for single-crystal PDIF-CN$_2$ FET on a BCB/SiO$_2$/Si-n$^{++}$ dielectric and substrate. **b**, charge carrier mobility vs. temperature for PDI8-CN$_2$, with and without contact resistance correction. These measurements are representative of our sample statistics,[5] i.e. more than 4 samples each one containing devices with different L (and constant W).

*Space-charge-limited current analysis*

In order to probe the bulk trap density in our system, we used the Space-Charged-Limited Current method. This method applied on two-terminal devices is based on the analysis of three different current regimes. The first region is governed by Ohm's law, where the current is proportional to the voltage applied. Then, the second is described by Child's law for solid, with a proportionality of V² to the current, where the end is characterized by a brutal increase of the current due to the filling of all trap levels at a certain voltage called trap-filled limit voltage, and mark the beginning of the third regime. The third region could be described by the Mott-Gurney Law, assuming that the current is only due to one type of carriers, without any effect of diffusion and with a field-independent mobility. To reach this last regime, a specific trap-filling voltage ($V_{TFL}$) has to be overcome, reflecting the voltage needed to fill all the charge traps in the material. According to Space-Charge-Limited Current theory, the bulk trap density ($N_t$) in a given semiconductor is proportional to that trap-filled limit voltage $V_{TFL}$ and its electrical permittivity, and can be extracted from the I-V curve by the following equation:

$$N_t = \frac{\varepsilon V_{TFL}}{eL^2}$$

Where $\varepsilon$ the dielectric constant of the material, e the elementary charge and L the distance between the electrodes.

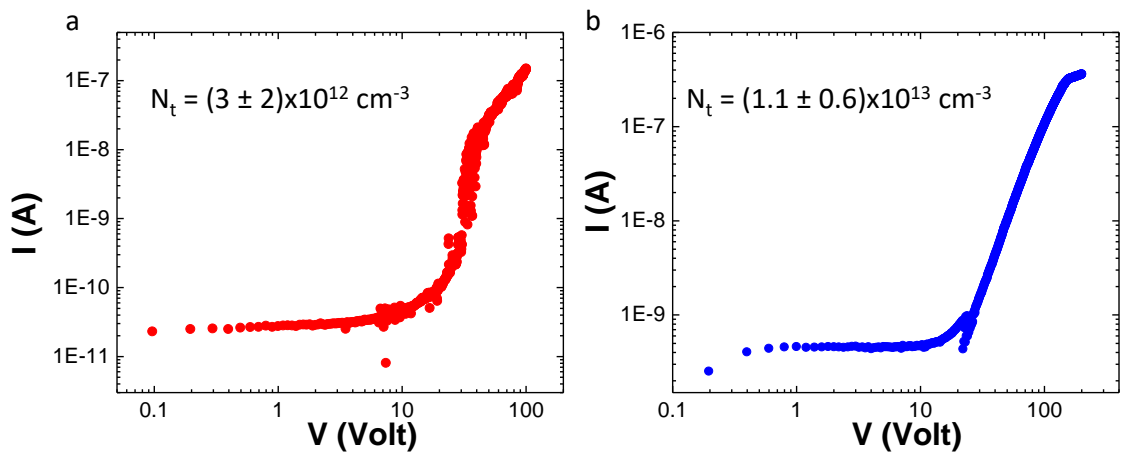

**Figure S17**: *Space-charge limited current analysis*. Current-voltage trace recorded on single crystals of **a**, PDI8-CN$_2$ and **b**, PDIF-CN$_2$. The first part of the curve is nearly ohmic, and it is followed by a quadratic regime following Child's law.[6,7] The bulk trap density $N_t$ was extracted over few samples at the voltage where the trap-filling transition is completed. $N_t$ was found to be roughly 3 times larger for PDIF-CN$_2$ than PDI8-CN$_2$ crystals

## 4. Temperature-dependent structural characterization

### a. X-Ray diffraction as a function of temperature

X-Ray diffraction measurements were performed on powder and single crystals at the X-ray diffraction beamline (XRD1) (Elettra-Sincrotrone Trieste S.C.p.A., Trieste, Italy).[8] Temperature dependent measurements were carried using an Oxford Cryostream 700 series (Oxford Cryosystems Ltd., Oxford, United Kingdom) from 100 K to 300 K. Data have been collected using a monochromatic wavelength of 0.700 Å (17.71 keV) and 200·200 µm$^2$ spot size, using a Pilatus 2M hybrid-pixel area detector (DECTRIS Ltd., Baden-Daettwil, Switzerland). The PDI8-CN$_2$ powder has been packed in borosilicate capillaries with a 300 µm diameter (10 µm wall thickness) or suspending crystals (scratched from silica support), in NHV Oil drops (Jena Bioscience, Jena, Germany) supported on kapton loops (MiTeGen, Ithaca, USA). Powder diffraction data have been collected in transmission mode, spinning the sample. Bi-dimensional powder patterns have been integrated using Fit2D program[9] after preliminary calibration of hardware setup, using a capillary filled with LaB$_6$ standard reference powder (NIST 660a). Randomly oriented single crystals of PDIF-CN$_2$ were dipped in NHV oil and mounted on the goniometer head with kapton loops. The diffraction data were indexed and integrated using XDS.[10] Scaling have been done using CCP4-Aimless code.[11,12] Fourier analysis and refinement were performed by the full-matrix least-squares methods based on F$^2$ implemented in SHELXL-2018/3,[13] starting from the coordinates already published.[14] Pictures were prepared using Mercury,[15] Ortep3 [16] and Pymol software.[17] Essential crystal and refinement data (Table S2) are reported below. The known triclinic crystalline form has been confirmed[18] and no phase transitions have been detected upon temperature ramping down to 100 K and up, back to room temperature (Figure S16). The crystal packing shows half PDIF-CN$_2$ in the crystallographic asymmetric unit (ASU – Figure S17, S18), since a crystallographic inversion center lays in the molecule baricenter. The molecule intrinsically lacks this element of symmetry therefore the cyano groups on the PDI bay appear disordered between two almost equally populated conformations. The same crystalline packing and PDI sidechains conformations are conserved in the whole temperature range explored: the molecular models show equivalent sidechains and PDI cores geometries and can be perfectly superimposed (R.M.S.D between models < 0.05 Å – Figure S19).

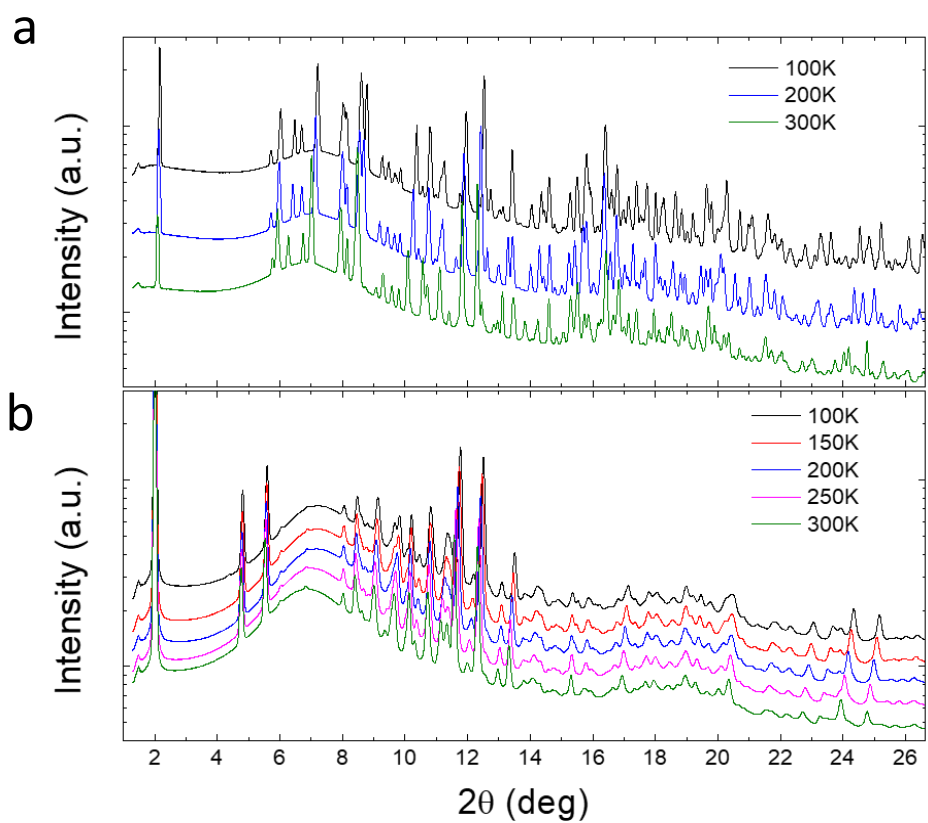

**Figure S18**: **a**, diffraction patterns of PDIF-CN$_2$ powder collected at different temperatures **b**, diffraction patterns of PDI8-CN$_2$ powder collected at different temperatures, where some Bragg peaks splitting and shifts were observed in the diffraction patterns during thermal treatment, indicating a small anisotropic contraction of the unit cell. The variation is smaller than that observed in PDIF-CN$_2$ crystals. No crystals of PDI8-CN$_2$, suitable for single crystal XRD structural determination, have been found.

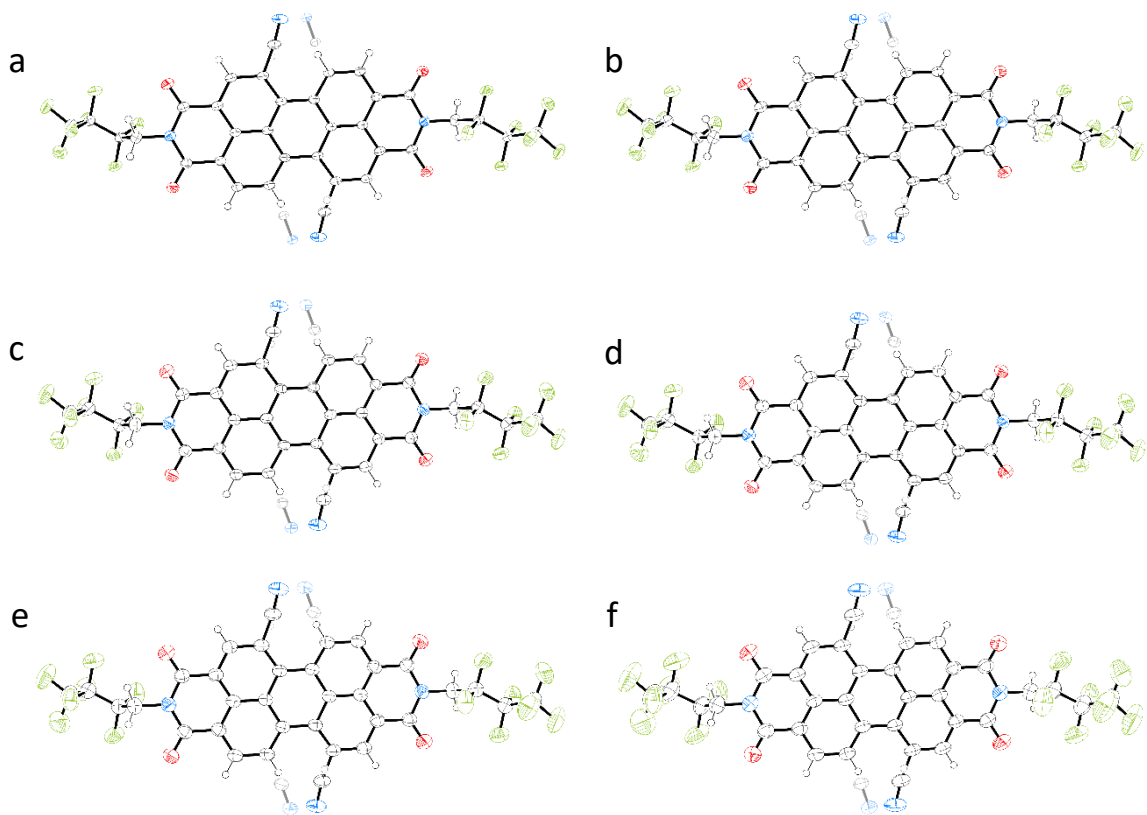

**Figure S19.** Ellipsoids representation of molecular structures (50% probability) of PDIF-CN$_2$ at **a,** 100 K, **b**, 150 K, **c**, 175 K, **d**, 200 K, **e**, 250 K and **f**, 300 K. Disordered CN groups conformations are greyed out for clarity.

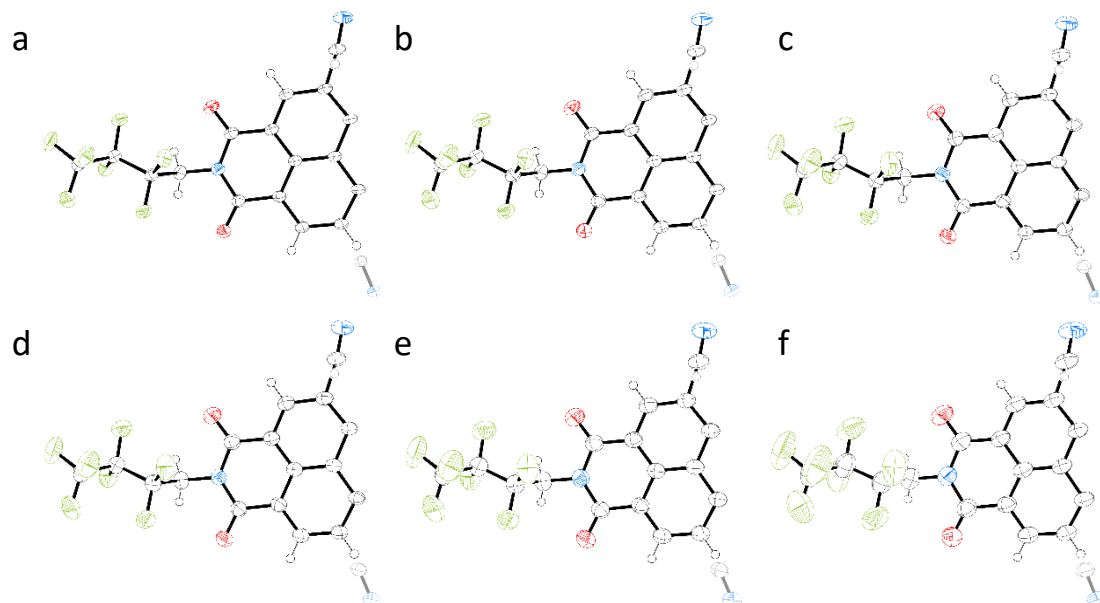

**Figure S20.** Ellipsoids representation of molecular structures (50% probability – ASUs contents) of PDIF-CN$_2$ at **a,** 100 K, **b**, 150 K, **c**, 175 K, **d**, 200 K, **e**, 250 K and **f**, 300 K. Disordered CN groups conformations are greyed out for clarity.

| PDIF-CN$_2$ | 100 K | 150 K | 175 K | 200 K | 250 K | 300 K |
|---|---|---|---|---|---|---|
| CCDC Number | 1848556 | 1848559 | 1848555 | 1848558 | 1848557 | 1848554 |
| Chemical Formula | C$_{34}$H$_{10}$F$_{14}$N$_4$O$_4$ | C$_{34}$H$_{10}$F$_{14}$N$_4$O$_4$ | C$_{34}$H$_{10}$F$_{14}$N$_4$O$_4$ | C$_{34}$H$_{10}$F$_{14}$N$_4$O$_4$ | C$_{34}$H$_{10}$F$_{14}$N$_4$O$_4$ | C$_{34}$H$_{10}$F$_{14}$N$_4$O$_4$ |
| Formula weight (g/mol) | 804.46 | 804.46 | 804.46 | 804.46 | 804.46 | 804.46 |
| Temperature (K) | 100 (2) | 150(2) | 175(2) | 200(2) | 250(2) | 300(2) |
| Wavelength (Å) | 0.700 | 0.700 | 0.700 | 0.700 | 0.700 | 0.700 |
| Crystal system | Triclinic | Triclinic | Triclinic | Triclinic | Triclinic | Triclinic |
| Space Group | $P$-1 | $P$-1 | $P$-1 | $P$-1 | $P$-1 | $P$-1 |
| Unit cell dimensions | $a$ = 5.222(1) Å | $a$ = 5.234(1) Å | $a$ = 5.238(1) Å | $a$ = 5.238(1) Å | $a$ = 5.259(1) Å | $a$ = 5.295(1) Å |
| | $b$ = 7.614(2) Å | $b$ = 7.626(2) Å | $b$ = 7.627(2) Å | $b$ = 7.626(2) Å | $b$ = 7.637(2) Å | $b$ = 7.612(2) Å |
| | $c$ = 18.753(4) Å | $c$ = 18.819(4) Å | $c$ = 18.858(4) Å | $c$ = 18.863(4) Å | $c$ = 18.994(4) Å | $c$ = 19.308(4) Å |
| | $\alpha$ = 92.85(3)° | $\alpha$ = 92.53(3)° | $\alpha$ = 92.33(3)° | $\alpha$ = 92.19(3)° | $\alpha$ = 91.75(3)° | $\alpha$ = 90.86(3)° |
| | $\beta$ = 95.13(3)° | $\beta$ = 95.31(3)° | $\beta$ = 95.37(3)° | $\beta$ = 95.47(3)° | $\beta$ = 95.54(3)° | $\beta$ = 95.38(3)° |
| | $\gamma$ = 104.71(3)° | $\gamma$ = 104.73(3)° | $\gamma$ = 104.73(3)° | $\gamma$ = 104.68(3)° | $\gamma$ = 104.84(3)° | $\gamma$ = 105.02(3)° |
| Volume (Å$^3$) | 716.3(3) | 721.6(3) | 723.8(3) | 724.0(3) | 732.7(3) | 747.7(3) |
| Z | 1 | 1 | 1 | 1 | 1 | 1 |
| Density (calculated) (g·cm$^{-3}$) | 1.865 | 1.851 | 1.846 | 1.845 | 1.823 | 1.787 |
| Absorption coefficient (mm$^{-1}$) | 0.176 | 0.175 | 0.174 | 0.174 | 0.172 | 0.169 |
| F(000) | 400 | 400 | 400 | 400 | 400 | 400 |
| Crystal size (mm$^3$) | 0.05 x 0.05 x 0.03 | 0.05 x 0.05 x 0.03 | 0.05 x 0.05 x 0.03 | 0.05 x 0.05 x 0.03 | 0.05 x 0.05 x 0.03 | 0.05 x 0.05 x 0.03 |
| Crystal habit | Thin red plates | Thin red plates | Thin red plates | Thin red plates | Thin red plates | Thin red plates |
| Theta range for data collection | 1.08° to 30.00° | 1.07° to 30.00° | 1.07° to 30.00° | 1.07° to 30.00° | 1.06° to 30.00° | 1.04° to 30.00° |
| Resolution (Å) | 0.70 | 0.70 | 0.70 | 0.70 | 0.70 | 0.70 |
| Index ranges | -7 ≤ h ≤ 7 | -7 ≤ h ≤ 7 | -7 ≤ h ≤ 7 | -7 ≤ h ≤ 7 | -7 ≤ h ≤ 7 | -7 ≤ h ≤ 7 |
| | -10 ≤ k ≤ 10 | -10 ≤ k ≤ 10 | -10 ≤ k ≤ 10 | -10 ≤ k ≤ 10 | -10 ≤ k ≤ 10 | -10 ≤ k ≤ 10 |
| | -26 ≤ l ≤ 26 | -26 ≤ l ≤ 26 | -26 ≤ l ≤ 26 | -26 ≤ l ≤ 26 | -27 ≤ l ≤ 27 | -27 ≤ l ≤ 27 |
| Reflections collected | 14275 | 14209 | 14584 | 14085 | 14811 | 19077 |
| Independent reflections (data with I>2σ(I)) | 4182 (3388) | 4216 (3587) | 4231 (3066) | 4243 (3370) | 4291 (3099) | 4420 (3098) |
| Data multiplicity (max resltn) | 3.28 (2.77) | 3.23 (2.75) | 3.30 (2.89) | 3.18 (2.67) | 3.30 (2.85) | 3.99 (3.22) |
| I/σ(I) (max resltn) | 21.92 (10.98) | 14.31 (10.21) | 12.44 (4.95) | 10.22 (4.96) | 14.61 (4.92) | 7.52 (2.88) |
| R$_{merge}$ (max resltn) | 0.0396 (0.1066) | 0.0397 (0.1103) | 0.0692 (0.1470) | 0.0586 (0.1328) | 0.0478 (0.1899) | 0.0867 (0.2014) |
| Data completeness (max resltn) | 96.0% (91.5%) | 95.9% (91.3%) | 95.9% (92.0%) | 95.9% (91.8%) | 95.6% (91.4%) | 95.3% (89.1%) |
| Refinement method | Full-matrix least-squares on F$^2$ | Full-matrix least-squares on F$^2$ | Full-matrix least-squares on F$^2$ | Full-matrix least-squares on F$^2$ | Full-matrix least-squares on F$^2$ | Full-matrix least-squares on F$^2$ |
| Data / restraints / parameters | 4182 / 0 / 273 | 4216 / 0 / 273 | 4231 / 0 / 273 | 4243 / 0 / 273 | 4291 / 0 / 273 | 4420 / 0 / 273 |
| Goodness-of-fit on F$^2$ | 1.003 | 1.044 | 1.040 | 1.011 | 1.076 | 1.089 |
| Final R indices [I>2σ(I)][a] | R$_1$ = 0.0508, wR$_2$ = 0.1397 | R$_1$ = 0.0538, wR$_2$ = 0.1425 | R$_1$ = 0.0539, wR$_2$ = 0.1527 | R$_1$ = 0.0487, wR$_2$ = 0.1279 | R$_1$ = 0.0549, wR$_2$ = 0.1609 | R$_1$ = 0.0861, wR$_2$ = 0.2708 |
| R indices (all data)[a] | R$_1$ = 0.0619, wR$_2$ = 0.1489 | R$_1$ = 0.0633, wR$_2$ = 0.1532 | R$_1$ = 0.0687, wR$_2$ = 0.1625 | R$_1$ = 0.0603, wR$_2$ = 0.1418 | R$_1$ = 0.0711, wR$_2$ = 0.176 | R$_1$ = 0.1030, wR$_2$ = 0.2970 |
| Largest diff. peak and hole (e·Å$^{-3}$) | 0.626 and -0.337 | 0.545 and -0.393 | 0.678 and -0.398 | 0.448 and -0.335 | 0.426 and -0.346 | 0.598 and -0.434 |
| R.M.S.D. from mean (e·Å$^{-3}$) | 0.080 | 0.071 | 0.082 | 0.063 | 0.058 | 0.087 |

[a] R$_1$ = Σ ||$F_o$|−|$F_c$|| / Σ |$F_o$|, wR$_2$ = {Σ [w($F_o^2$ − $F_c^2$)$^2$] / Σ [w($F_o^2$)$^2$]}$^{½}$

**Table S2.** Crystallographic data and refinement details for PDIF-CN$_2$ at 100 K, 150 K, 175 K, 200 K, 250 K and 300 K.

## b. Low-temperature Selected-Area Electron Diffraction characterization

Transmission Electron Microscopy (TEM) characterization was performed on PDI8-CN$_2$ and PDIF-CN$_2$ crystals following a drop-casting deposition on TEM grids. The measurements were performed with an FEI ECNAI F20 microscope operating at 120 keV at 77 K. The TEM images were taken in phase contrast mode and Selected Area Electron Diffraction (SAED).

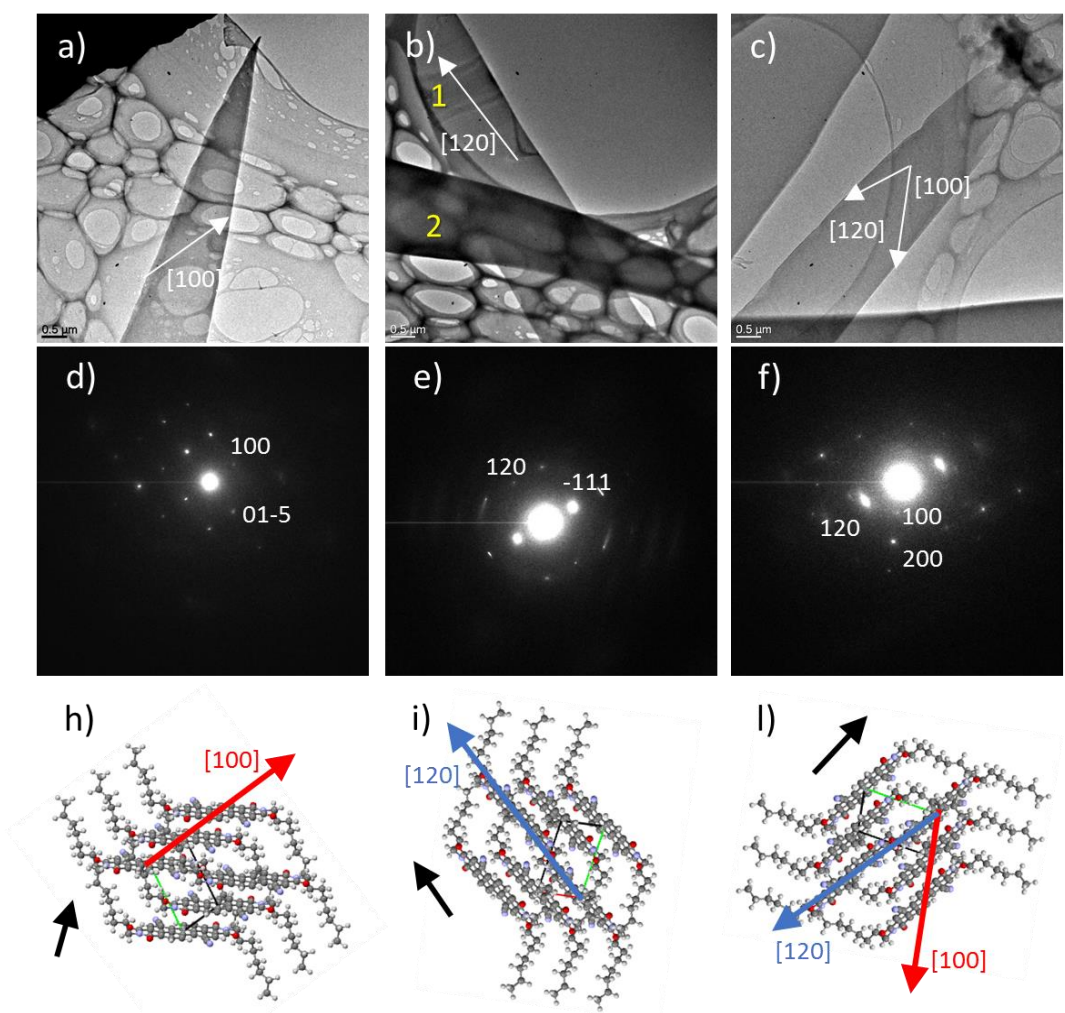

**Figure S21**: (a-c) TEM images and the corresponding (d-f) SAED patterns of different PDI8-CN$_2$ crystals. The white arrows in TEM images indicate the crystallographic directions detected by SAED. (h-l) Sketch of the molecular arrangement in a crystal at the surface as found through SAED/TEM analysis (top of view). Black arrows indicate the crystal direction.

TEM images were collected on several PDI8-CN$_2$ crystals, some of them are reported in Figure S20. The corresponding SAED patterns confirm the single crystalline nature of the crystals and show different zone axis related to the PDI8-CN$_2$ crystal structure.[18] In particular, they reveal that the crystals may grow along different crystallographic directions. For instance, Figure S20a shows a crystal whose unit cell is aligned such that the π-π stacking direction is almost parallel to the crystal direction (see sketch in Figure S20h). In contrast, the crystal "1" in Figure S20b, as well as the fiber in Figure S20c, exhibit the π-π stacking almost perpendicular to the crystal direction. The measured crystals were ca. 1-μm thick. Thicker crystals (e.g. number "2" in Figure 20b) were deformed by the radiation damage and/or the local heating due to the high electron absorption.

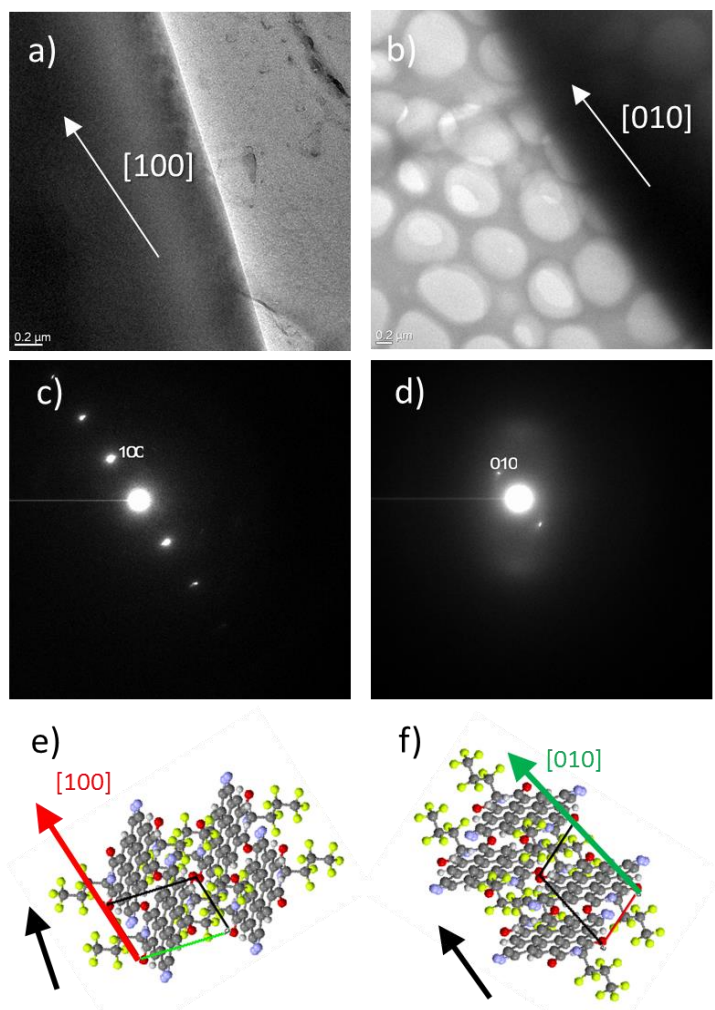

**Figure S22**: TEM (a-c) and the corresponding SAED (d-f) of different PDIF-CN$_2$ crystals. White arrows in TEM images indicate the crystallographic directions detected by SAED. c) Sketch of the molecular arrangement in a crystal at the surface as found through SAED/TEM analysis (top of view). Black arrows indicate the crystal direction.

The examined PDIF-CN$_2$ crystals were thicker than those of PDI8-CN$_2$ employed for the measurement. Therefore, TEM images were collected at the fiber edge. Figure S21 shows the TEM and the corresponding SAED images pointing out the different crystallographic growth directions also for PDIF-CN$_2$ crystals. In this case the crystal can grow along the [100] (Figure S21a) or [010] (Figure S21b) direction. In both samples, there was a π-π interaction along the crystals direction.

**c.     Magic-angle spinning (MAS) solid-state NMR spectra as a function of temperature**

Fluorine Solid state NMR experiments were done on an AVANCE 500 MHz wide bore spectrometer (BrukerTM) operating at a frequency of 470.45 MHz for 19F and equipped with a triple resonance MAS probe designed for 2.5 mm o.d. zirconia rotors (closed with Kel-F caps) and a BCU extreme for temperature regulation.  In order to properly simulate spectra undistorted, lineshapes are needed, therefore all spectra were acquired with the original Hahn's echo sequence[19] (without 1H decoupling). This echo was synchronized with the MAS rotation and set equal to two rotation periods for the different MAS speeds (ranging from 15 to 26 kHz). FIDs were sampled over 8192 time-domain points spaced by 1 µs Dwell time, leading to a 122.07 Hz and 500 kHz (1062.8142 ppm) spectral resolution and width respectively. 16 scans were added for full phase cycle completion and noise averaging. A Lorentzian line broadening of 150 Hz was applied prior to Fourier transformation that was done on 16384 points (zero fill to 16 K). The referencing was done by setting 19F PTFE signal at -122 ppm (room temp).

The spectra were treated with Topspin software. Through the deconvolution process of each peak, the chemical shift, the chemical shift anisotropy (CSA) and the integral can be extracted as relevant information. The spectrum generated by the deconvolution process is compared to the measured one with an indication of the percentage of overlapping between them. For all the analysis, the lowest value of overlapping reached is 93.86%, for an average of 96.1%.

From that fitting, 6 chemically equivalent atoms are found, due to the position on the chain and the interaction with the atoms of the neighbouring molecules. In case of phase transition, since the environment of the atoms should change, the CSA should change as well or additional peaks would appear. However, no particular changes can be observed at low temperature. The fluctuations observed after 220 K are coming from the reduction of the spin rate to 15.140 kHz, which increase the incertitude of the deconvolution.

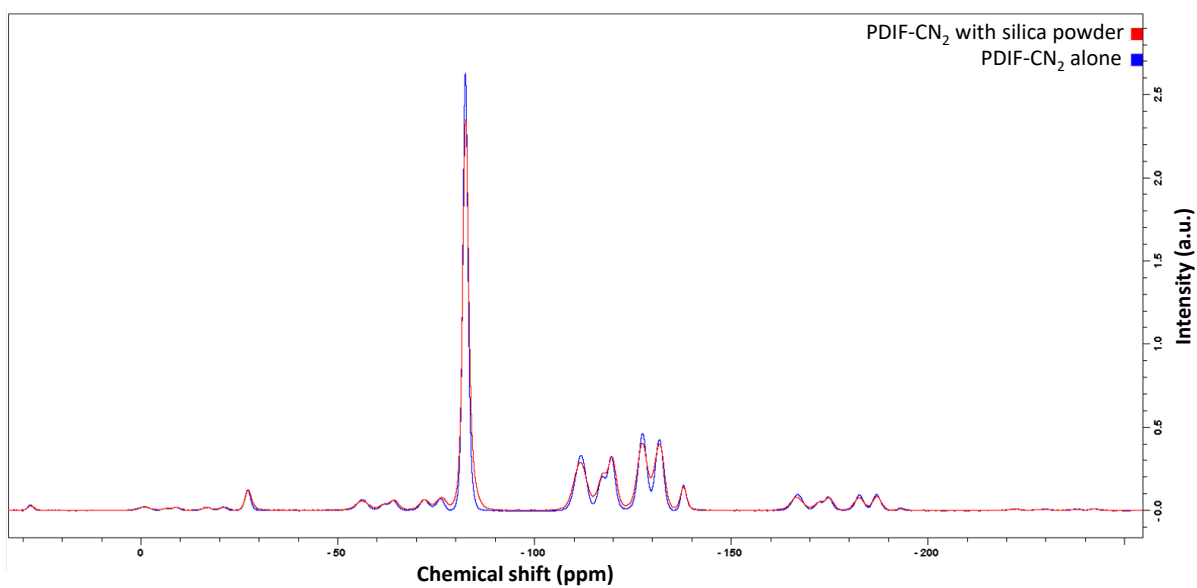

**Figure S23**: F[19] Magic-angle spinning (MAS) solid-state NMR spectrum[20] of PDIF-CN$_2$ at 300 K with a spinning rate of 20 kHz, and blended with silica powder. An electromagnetic induction can be generated in electrical conductor and semiconductor when they are in movement under a magnetic field. This induction can perturb the probe of the NMR. By isolating the grains with an inert material, like silica powder, these interactions with the probe are minimized.[21] We can observe that the signal of PDIF-CN$_2$ is not altered by the presence of the silica powder that just leads to a small reduction in the signal intensity.

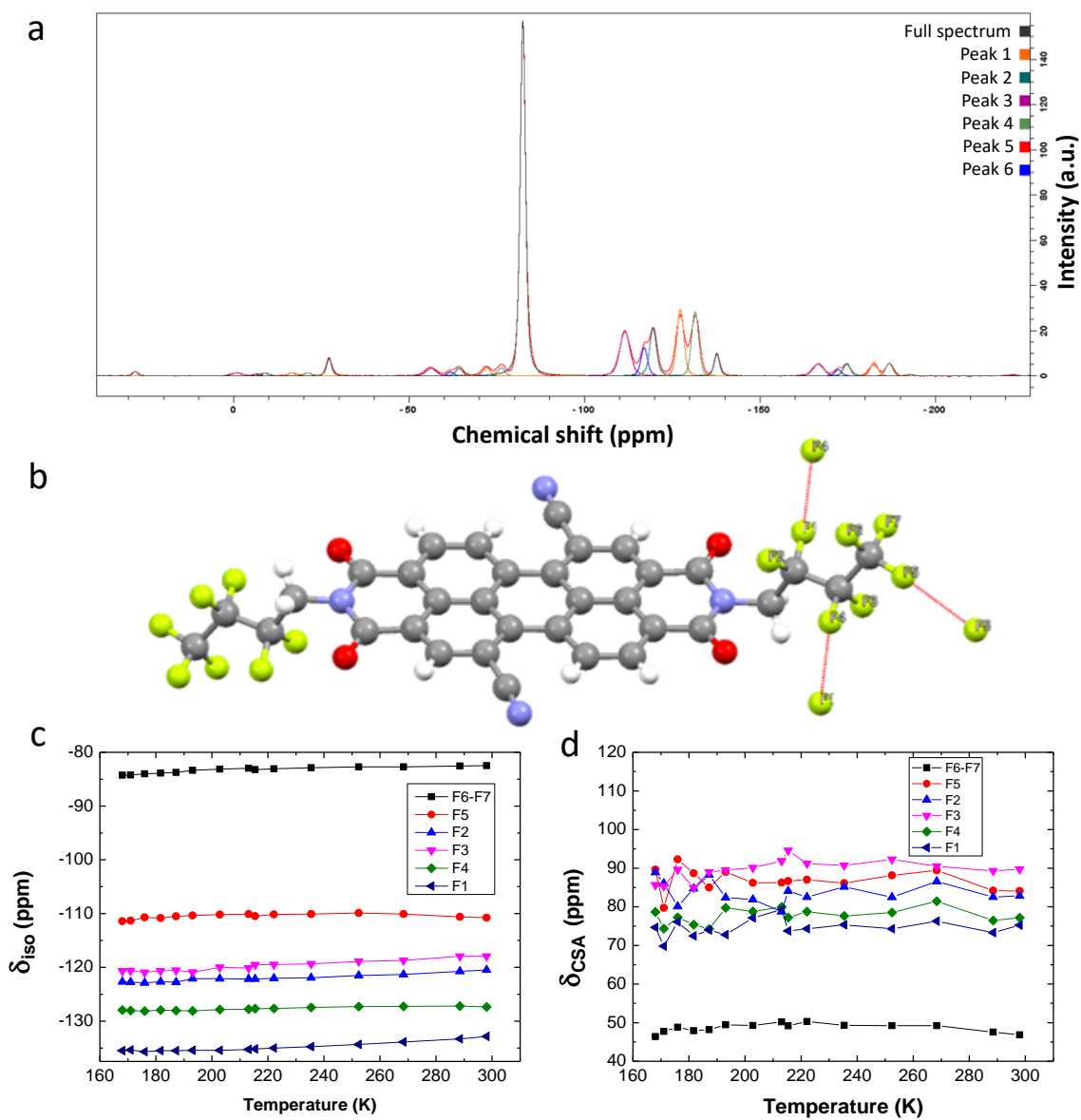

**Figure S24: a**, F[19] Magic-angle spinning (MAS) solid-state NMR spectrum[20] of PDIF-CN$_2$ at 298 K with a spinning rate of 20 kHz, showing the presence of 6 chemically equivalent peaks. **b**, X-ray structure of PDIF-CN$_2$ with specific labels on the fluorine atoms, **c**, and **d**, evolution of the chemical shift and the chemical shift anisotropy (CSA) respectively, of the 6 groups of atoms chemically equivalent, between 298 K and 167 K, extracted from the spectra. The constant values of both chemical shifts indicate that the environment of the fluorine atoms is not changing when cooling down. Any phase transition would appear in the change of the chemical shift anisotropy or by the appearance/disappearance of peaks. The increase of the noise is coming from the reduction of the spinning rate below 220 K to 15.140 kHz.

## 5. Atomic Force/ Kelvin Probe Force microscopy characterization of single crystals

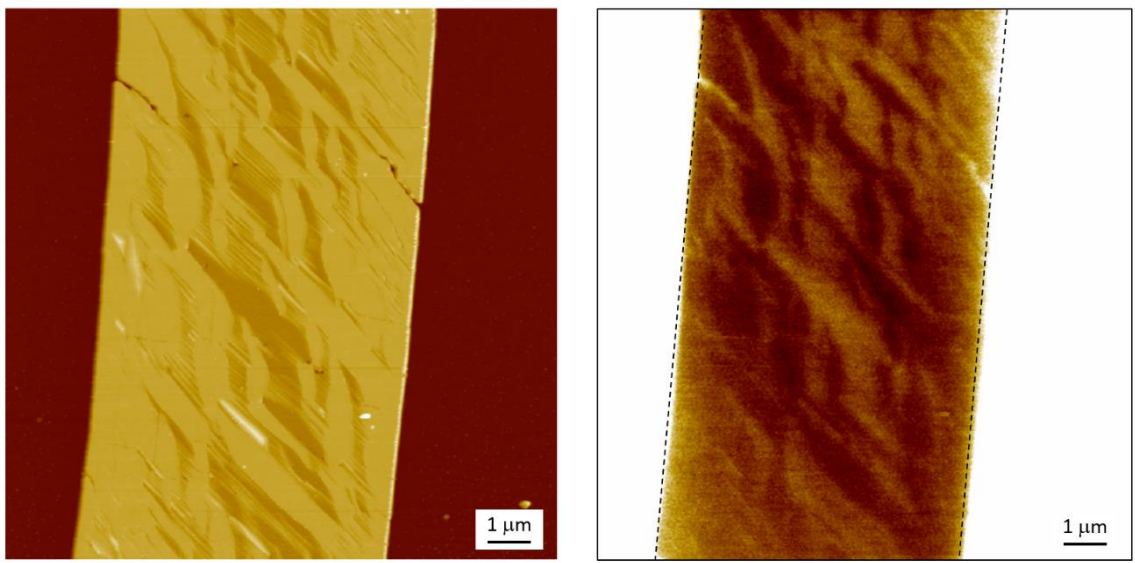

**Figure S25**: (*Left*) AFM and (*Right*) KPFM image of a single PDIF-CN$_2$ crystal [scales: 200 nm, 180 meV, respectively]

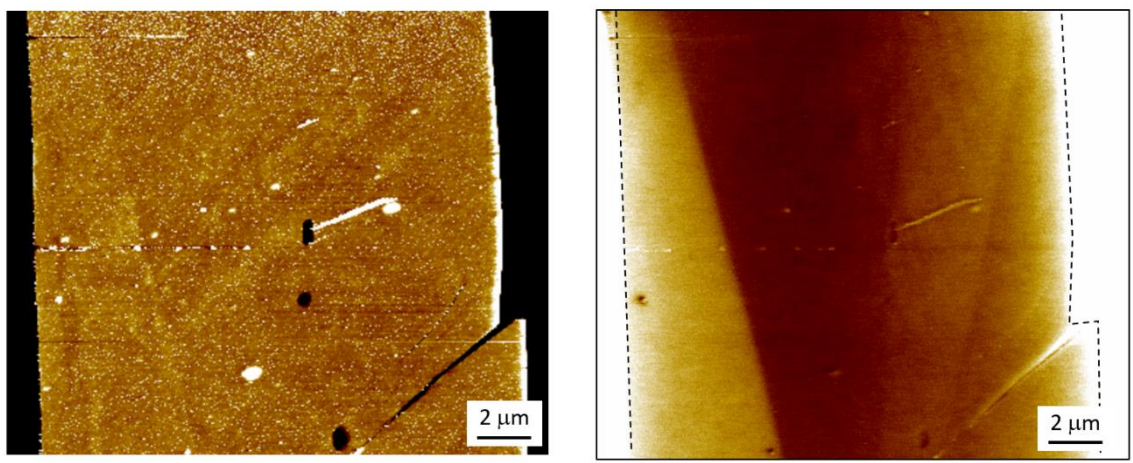

**Figure S26:** (*Left*) AFM and (*Right*) KPFM image of a single PDI8-CN$_2$ crystal [scales: 7 nm, 180 meV, respectively]

AFM measurements revealed that all the electrically probed crystals were 100 to 200 nm thick. SPM measurements were carried out in ambient conditions by employing a commercial digital microscope Multimode 8 (Bruker) using Pt/Ir-coated cantilever silicon tips (Bruker SCM-PIT-V2, k = 2.8 N/m) with oscillating frequencies of 75 kHz. AFM and KPFM images were acquired during the same measurement (ambient conditions). A topographic line scan is first acquired by AFM operating in tapping mode, and then that same line is rescanned in lift mode with the tip raised to a lift height of 100 nm using the amplitude modulation (AM) mode. Raw AFM data were treated by using histogram-flattening procedures[22] to remove the experimental artifacts due to the piezoscanner. Step edges at the surface of single crystals were found to trap electrons in both p-type and n-type organic semiconductor crystals. In particular, charge carrier mobilities would decrease with increasing step density. Interestingly, the same study also showed that the thickness of the crystals can influence the type of transport, which can go from thermally activated, in very thick crystals (> 30-µm thick), to band-like transport, in thinner crystals (ca. 5-µm thick). This former experimental observation agrees well with the suggested relationship of proportionality between thickness and density of step edges. More specifically, in the case of PDIF-$CN_2$, the authors demonstrated that charge carrier mobility increases with decreasing the step density. Our AFM measurements carried out on the surface of the very same crystals that were electrically probed in the devices, revealed a multi-layered structure for both PDIF-$CN_2$ and PDI8-$CN_2$ crystals with typical thicknesses falling within 100 nm. For both derivatives, no step edges were measured on the crystals' surface as revealed by KPFM. In this regard, our experimental observation is in line with what suggested by T. He et al. in that our thin crystals represent the lower limit of the suggested investigation in which the level of static disorder is minimal thanks to the absence of step edges at which electrons could be trapped.

## 6. Modeling and simulation

### a. Molecular Dynamics simulations

Molecular Dynamics (MD) simulations are run using a reparametrized version of the Dreiding force field[23] where the atomic charges of the PDI core have been obtained by fitting the Electrostatic Potential (ESP)[24] as calculated at the Density Functional level of theory (DFT) using the PBE0 functional. We have built a supercell made of 64 molecules (8 x 4 x 2) starting from the experimental unit cell. We have first run a NVT equilibration dynamics of 100 ps at 298 K, considering a timestep of 1 fs and Ewald summation for the calculation of electrostatic interactions together with the use of the Nosé-Hoover thermostat. Then, we have run a NVT production dynamics of 5 ps at 298 K, considering a timestep of 1 fs, Ewald summation and the Nosé-Hoover thermostat. Configurations of the system are saved every 5 fs. The simulation time and the time interval between saved configurations of the production dynamics are chosen to allow for accurate sampling of low-frequency phonons modes. MD simulations have been performed with the Materials Studio (MS) 6.0 code.

| | | $\langle d_a \rangle$ (Å) | $\sigma_{d_a}$ (Å) | $\langle d_b \rangle$ (Å) | $\sigma_{d_b}$ (Å) | $\langle d_c \rangle$ (Å) | $\sigma_{d_c}$ (Å) |
|---|---|---|---|---|---|---|---|
| **PDIF-CN$_2$** | π-π | 5.232 | 0.138 | 0.082 | 0.061 | 0.110 | 0.082 |
| | π-Edge | 5.232 | 0.143 | 7.638 | 0.107 | 0.100 | 0.078 |
| **PDI8-CN$_2$** | π-π | 5.019 | 0.159 | 0.153 | 0.109 | 0.149 | 0.115 |
| | π-Edge | 5.019 | 0.181 | 8.841 | 0.181 | 0.144 | 0.111 |

**Table S3**: Analysis in terms of molecular displacements along the crystalline axes, recorded along the molecular dynamics trajectory for the π-π and π-edge directions for both PDIF-CN$_2$ and PDI8-CN$_2$. The standard deviations of the displacements along the different molecular axis are highlighted in colour. The largest molecular displacements occur along the π stacking direction. PDI8-CN$_2$ displacements exhibit larger standard deviation compared to PDIF-CN$_2$.

### b. Microelectrostatic calculations

We have computed the contribution from intermolecular electrostatic interactions between polarizable model by means of classical atomistic microelectrostatic calculations performed with the MESCAL code.[25] These calculations have two main parameters as inputs, namely the polarizability tensor and the atomic ESP charges of the conjugated cores that are obtained at the DFT level using the PBE0 functional with the 6-311G basis set. The polarization energy (P) obtained from self-consistent microelectrostatic calculations can be decomposed as the sum of two contributions, electrostatics (S) and induction energy (D), as detailed in a previous work.[24] Calculations were performed on spherical molecular clusters of 40 Å radius, which ensure converged polarization energy differences between neighboring sites.

In Figure S26, we have reported the distributions of polarization, electrostatic and induction energies for both PDI derivatives. Even though the S and D distributions have different averages and standard deviations values because of the different packings of the PDI derivatives, it is interesting to notice that distributions of the total polarization energies P are very similar and, most importantly, characterized by almost the same standard deviation. The fluctuations in polarization energy are hence not the expected to be the decisive parameter that differentiate the electron transport properties of PDIF-$CN_2$ and PDI8-$CN_2$.

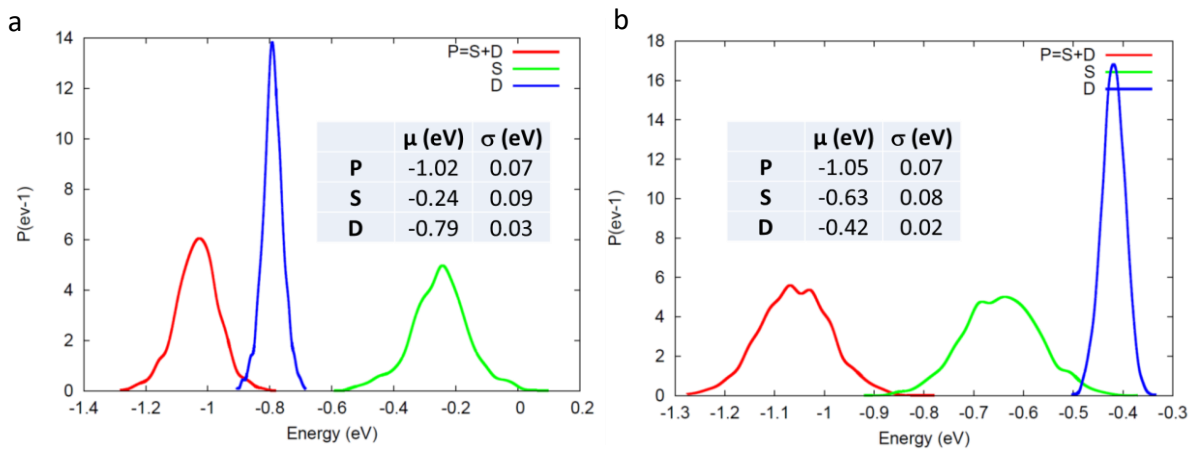

**Figure S27**: Distribution of polarization (P) as a sum of electrostatic (S) and induction (D) energies **a**, for PDI8-CN$_2$ and **b**, for PDIF-CN$_2$. The electrostatic contribution involves the interaction between permanent charges at atomic site while the induction contribution involves the interaction between the permanent charges and the induced dipoles.

### c. Transfer integrals calculations

The transfer integral has been calculated within the dimer *fragment approach* as implemented in the Amsterdam Density Functional (ADF) package[26] for nearest-neighbours PDI dimers along the pi-pi and pi-edge directions. In this approach, the orbitals of the dimer are expressed as linear combination of molecular orbitals of the fragments that are obtained by solving the Kohn-Sham equations. Especially, the site energies, $\varepsilon_1$ and $\varepsilon_2$, and the transfer integrals $t_{12}$, are obtained by computing the following matrix elements:

$$\varepsilon_i = \langle \varphi_i | \hat{H} | \varphi_i \rangle$$
$$t_{ij} = \langle \varphi_i | \hat{H} | \varphi_j \rangle$$

Where $\varphi_i$ and $\varphi_j$ correspond to the HOMO/LUMO orbitals of the isolated molecules (*i.e.* fragments). The transfer integral $t_{12}$ has been evaluated at density functional theory (DFT) level using the PBE functional[27] with a Double Zeta basis set. However, due to the non-orthogonality of the fragment orbital basis set, the transfer integral value is not uniquely defined and depends on the definition of the energy of the origin.[28] The problem is solved by applying a Löwdin transformation to the initial electronic Hamiltonian resulting in the following expression of the transfer integral:

$$\tilde{t}_{12} = \frac{t_{12} - (\varepsilon_1 + \varepsilon_2)S_{12}}{1 - S_{12}}$$

Where the parameter $S_{12}$ represents the orbitals overlap.

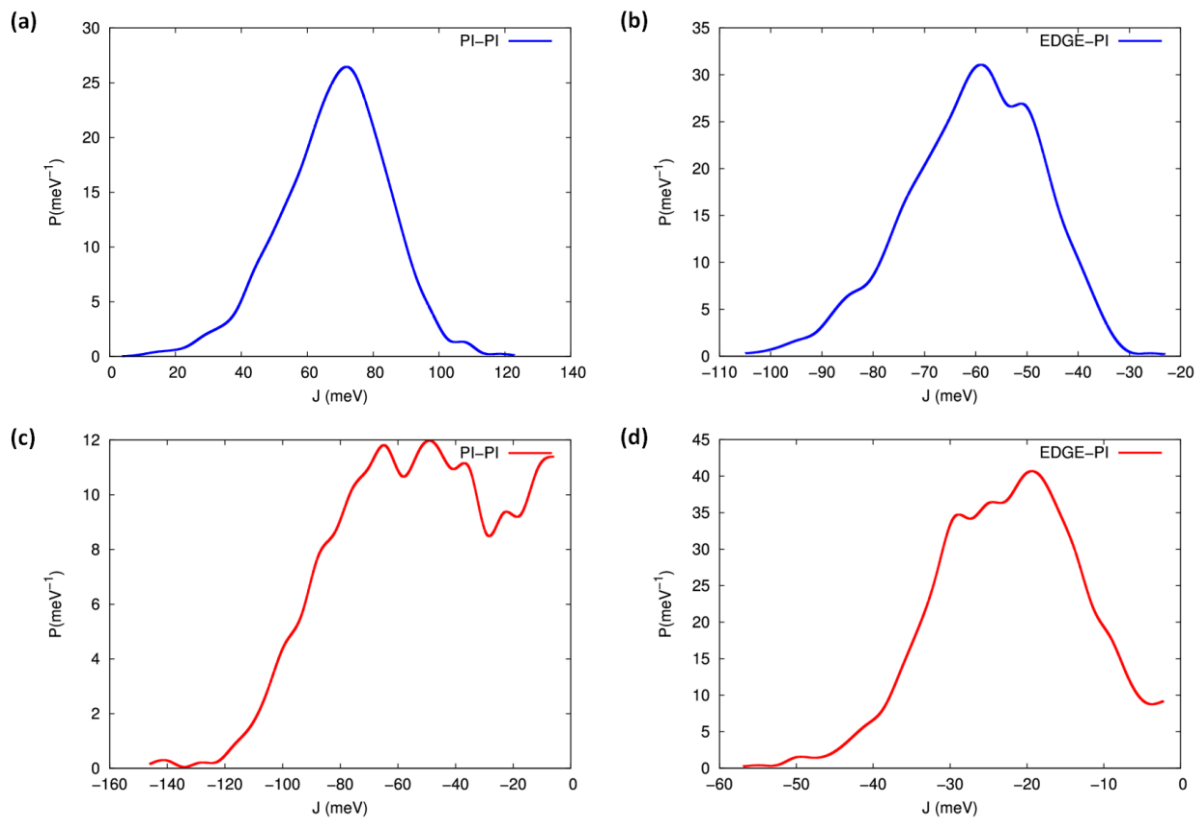

**Figure S28:** Histograms of electron transfer integrals for **a**, the π-π interaction and **b**, the π-edge interaction in PDIF-CN$_2$ structure, **c**, and **d**, for PDI8-CN$_2$.

|  | π-π | | π-edge | |
|---|---|---|---|---|
|  | μ (meV) | σ (meV) | μ (meV) | σ (meV) |
| **PDIF-CN2** | 70.8 | 16.4 | -58.7 | 13.2 |
| **PD8-CN2** | -48.5 | 28.8 | -21.0 | 9.6 |

**Table S4:** Averaged transfer integrals along π-π and π-edge directions for PDIF-CN$_2$ and PDI8-CN$_2$.

### d. Intermolecular vibrational modes

The simulated autocorrelation function of the polarization energy P at a time-lag $\tau_i$ is defined as:

$$\langle P(0)P(\tau_i)\rangle = \frac{1}{N(M-m-1)} \sum_n^N \sum_{m=0}^{M-m-1} P(m)P(i+m)$$

Where the sum over n (m) goes over the number of molecules in the unit cell (the snapshots of the MD trajectory)

Similarly, the simulated autocorrelation function of the transfer integrals J at a time-lag $\tau_i$ is defined as:

$$\langle J(0)J(\tau_i)\rangle = \frac{1}{N(M-m-1)} \sum_n^N \sum_{m=0}^{M-m-1} J(m)J(i+m)$$

The expression of the electron-phonon coupling constants $\nu_j$ is the following one:

$$\nu_j = \sqrt{\frac{\beta \hbar \Omega_j}{\pi} G(\Omega_j)}$$

Where $G(\Omega_j)$ is the Fourier transform of the autocorrelation of either the site energies or the transfer integrals along the π- π and the π -edge directions, $\Omega_j$ corresponds to a $\beta = \frac{1}{k_B T}$

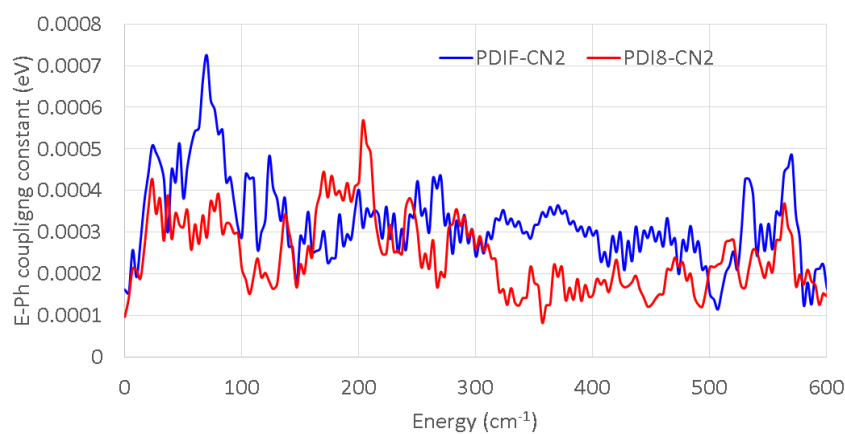

**Figure S29:** (Left panel) Electron-phonon coupling spectrum of the time-dependent transfer integrals along the edge-pi direction in crystals of PDI8-CN$_2$ (in red) and PDIF-CN$_2$ (in blue), respectively (T = T$_{amb}$). The lateral table contains, for each PDI derivative, the range of the calculated energy peak values and their related nature of displacement.

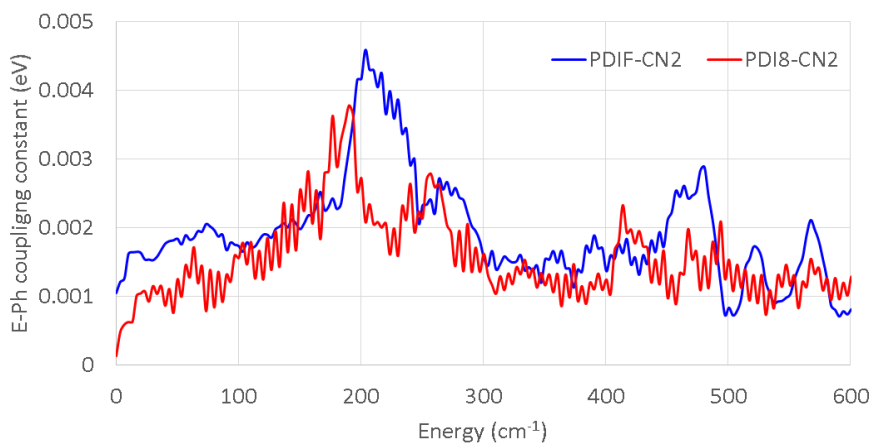

**Figure S30:** (Left panel) Electron-phonon coupling spectrum of the time-dependent site energies in crystals of PDI8-CN$_2$ (in red) and PDIF-CN$_2$ (in blue), respectively (T = T$_{amb}$). The lateral table contains, for each PDI derivative, the range of the calculated energy peak values and their related nature of displacement.

### e. Localization length and tight binding model

We have fed a tight-binding model with the polarization energies and the transfer integrals calculated along the MD trajectory considering one electronic level per molecular site and intersite electronic couplings along both the pi-pi and pi-edge directions so that the Hamiltonian (*H*) of a system made of N molecules writes as:

$$H = \sum_{i=1}^{N} P_i |i\rangle\langle i| + \sum_{i \neq j} J_{ij} |i\rangle\langle j| \text{ avec } |i\rangle \equiv |LUMO_i\rangle$$

Where $P_i$ is the polarization energy of molecule *i* and $J_{ij}$ is the transfer integral between molecule *i* and molecule *j*.

Diagonalizing *H* allows to access the adiabatic states of the system that have a certain degree of delocalization over the different molecules of the system. These states are linear combination of the LUMOs of the isolated molecules:

$$|\psi_m\rangle = \sum_{i=1}^{N} c_{im} |i\rangle$$

The localization length or Inverse participation ratio (IPR) for state m is calculated as:

$$IPR_m = \left( \sum_{i=1}^{N} c_{im}^4 \right)^{-1}$$

$IPR_m$ is calculated within the a-b plane considering 32 molecules since the transfer integrals along the c axis are negligible. We calculated the $IPR_m$ for 100 snapshots along a molecular dynamics trajectory. Thermalized IPR are calculated considering that the electron distributes following a Boltzmann distribution:

$$IPR_{TH} = \frac{\sum_i^N IPR_i \, exp(-(E_i - E_0)/k_B T)}{\sum_i^N exp(-(E_i - E_0)/k_B T)}$$

With $E_0$ the lowest energy adiabatic energy state (i.e. ground state).

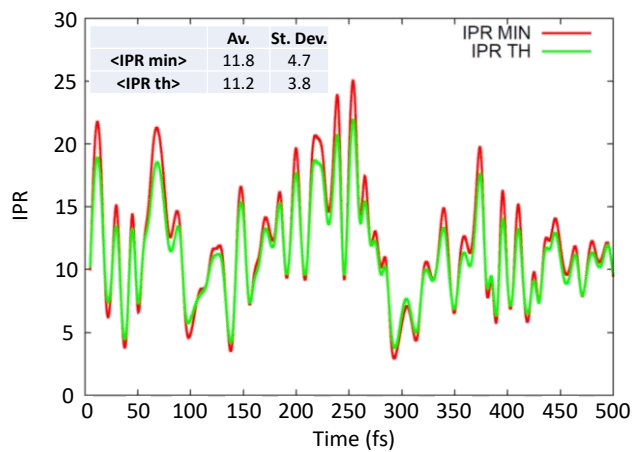

**Figure S31**: Time evolution of the localization length for the ground (IPR MIN) state of PDIF-$CN_2$ of the tight binding model considering the site energy, compared to the Boltzmann-averaged localization length (IPR TH). Its spread corresponds to the thermal disorder and involves a localization of the different states. Nevertheless, the localization length is larger than 10 molecules, which remains large compared to the size of the system (32 molecules).

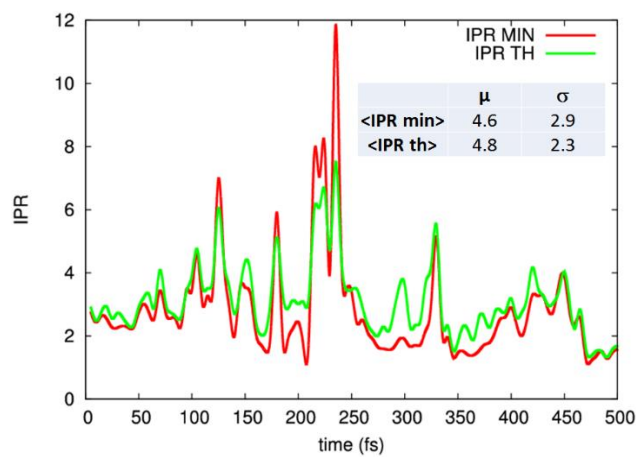

**Figure S32:** Time evolution of the localization length for the ground (IPR MIN) state of PDI8-$CN_2$ of the tight binding model considering the site energy, compared to the Boltzmann-averaged localization length (IPR TH). Both IPR MIN and IPR TH are similar, the site energies effect consideration is even stronger with a localization length largely reduced around 4 molecules, which is much smaller than for PDIF-$CN_2$.

### 7. Temperature-dependent µ-Raman spectroscopy on single crystals

Raman spectra were recorded with a Renishaw inVia confocal Raman microscope combined with a MS20 Encoded stage 100 nm allowing photoluminescence measurements with an excitation at 532 nm maintained at a power excitation below 1 mW. The atmosphere and the temperature were controlled with a Linkam THMSG600 stage. The measurements were carried out with a 50x LW lens affording a beam spot of 2 μm.

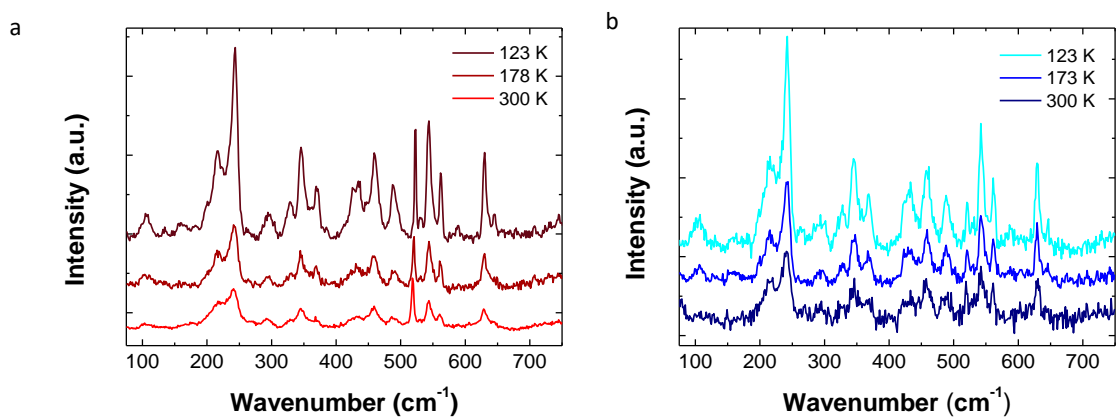

**Figure S33:** Raman spectra of **a**, PDIF-CN$_2$ and **b**, PDI8-CN$_2$, at different temperatures. The baselines have been corrected by subtracting the photoluminescence signal for the corresponding molecule and temperature.

## 8. Temperature-dependent Inelastic Neutron Scattering

The temperature-dependent inelastic neutron scattering (INS) measurements were performed using the direct geometry cold neutron, time-focusing time-of-flight spectrometer IN6 at the Institut Laue-Langevin (ILL) (Grenoble, France). About 250 mg of powder samples of PDIF-CN$_2$ and PDI8-CN$_2$, prepared as described above, were sealed inside thin flat Aluminum holder that was fixed to the sample stick of a cryofurnace. Data were collected at 150 K, 200 K, 300 K and 420 K for PDIF-CN$_2$, and at 200 K, 300 K and 420 K for PDI8-CN$_2$. The INS spectra were collected in the up-scattering regime (neutron energy-gain mode) using the high-resolution mode and a neutron incident wavelength $\lambda_i$=4.14 Å ($E_i$ = 4.77 meV), corresponding to a maximum Q ~ 2.6 Å$^{-1}$ on IN6, and offering a good resolution within the considered dynamical range for the anti-Stokes data. Standard corrections including detector efficiency calibration and background subtraction were performed. The data analysis was done using ILL procedures and software tools to extract the *Q*-averaged, one-phonon, generalized density of states (GDOS) from the INS measurements.[29–31] The GDOS is formulated as:

$$g^{(n)}(E) = B \sum_i \left\{\frac{4\pi b_i^2}{m_i}\right\} x_i g_i(E) \tag{1}$$

where $B$ is a normalization constant and $b_i$, $m_i$, $x_i$, and $g_i(E)$ are, respectively, the neutron scattering length, mass, atomic fraction, and partial density of states of the i$^{th}$ atom in the unit cell. The weighting factors $\frac{4\pi b_i^2}{m_i}$ for various atoms in the units of barns.amu$^{-1}$ are[32] : H: 81.37, C: 0.46, N: 0.82, O: 0.26, and F: 0.21.

The present cold-neutron INS spectra, in terms of the one-phonon generalised density of states (GDOS), provides the most low-energy information for all samples, with a limited resolution at high energy as compared to the other techniques. This resolution can be improved by collecting INS spectra down to base temperatures (to reduce the broadening due to the Debye-Waller factor), using a hot-neutron spectrometer, at the price of worsening the low-energy precision. In summary, INS offers a better alternative when it comes to explore both the external (phonons or lattice dynamics) and internal (molecular vibrations) modes as INS covers the whole Brillouin zone without the constraint of selection rules. Neutrons interact with samples nuclei and the resulting intensities are weighted by the ratio σ/M where σ and M are the nuclear cross section and M the mass of the scatterer, respectively. Therefore, in hydrogenated materials, the vibrational spectrum is going to be dominated by the dynamical density of freedoms (d.o.f.) involving hydrogen motions (the lightest element with the highest cross section).

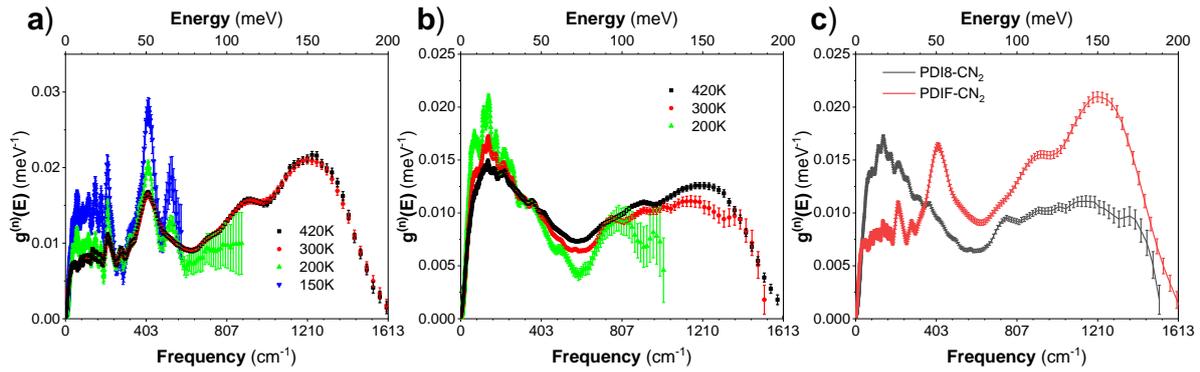

**Figure S34:** The temperature evolution of the inelastic neutron scattering spectra in terms of one-phonon GDOS of **a**, PDIF-CN$_2$ and **b**, PDI8-CN$_2$ on the full range of temperature. **c**, comparison of PDIF-CN$_2$ and PDI8-CN$_2$ INS spectra collected at 300 K.

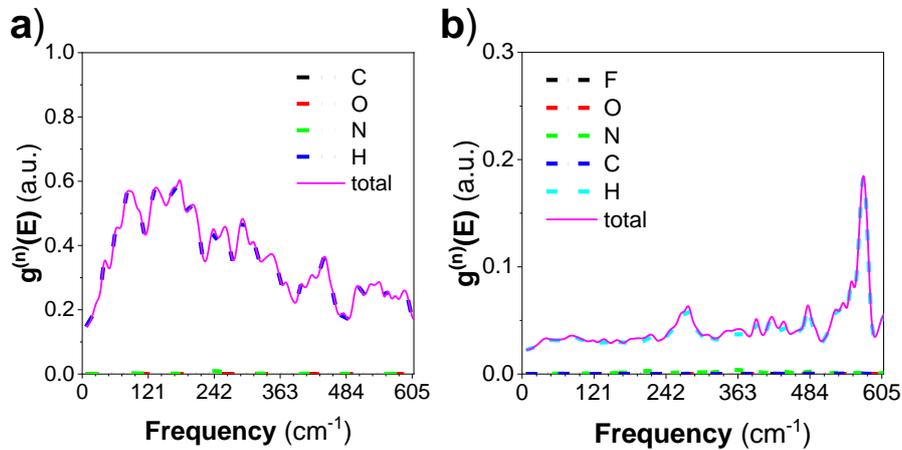

**Figure S35:** Simulated partial (H, C, N, O and F) and total GDOS of **a**, PDI8-CN$_2$ and **b**, PDIF-CN$_2$

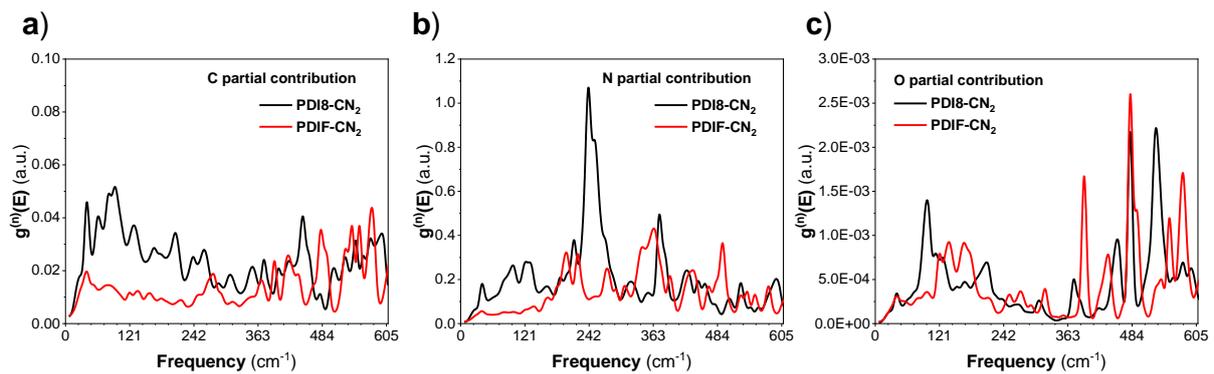

**Figure S36:** Comparison of the simulated partial (C, N and O) GDOS of PDI8-CN$_2$ and PDIF-CN$_2$

## 9. Electron energy loss spectroscopy and ultraviolet photoelectron spectroscopy measurements

The as received samples were mounted on a sample holder and inserted in the analysis chamber (base pressure $1.10^{-10}$ mBar) for the analysis. Spectra were collected with a double pass cylindrical mirror analyzer (PHI-15-255GAR) driven at constant pass energy. Energy resolution was set to 0.1 eV. UPS was taken using a He discharge lamp from VG (photon energy 21.2 eV). The EELS spectra were collected using an electron beam at 160 eV and at a current of 10 nA. EELS measurements of the optical band gap ($E_{opt}$) of the crystals of both compounds provided nearly identical values for PDI8-CN$_2$ and PDIF-CN$_2$ (Fig. S36a). While $E_{opt}$ and presumably the electronic band gap (HOMO-LUMO) of PDI8-CN$_2$ and PDIF-CN$_2$ are nearly equivalent, their respective ionization energies (Fig. S36b and Table S5) were found to be offset by as much as 0.70 eV (amounting to 7.9 eV and 7.4 eV, for PDIF-CN$_2$ and PDI8-CN$_2$, respectively). Considering the measured energy difference between $E_F$ and HOMO level (Fig. S36c), it appears clear that electron injection is certainly more favorable in PDIF-CN$_2$ thanks to the reduction of the energy barrier between Au work function and its LUMO.

| Parameter | PDIF-CN$_2$ | PDI8-CN$_2$ |
|---|---|---|
| Work function of Si/SiO$_2$/PDI $\Phi$ (eV) | 5.20 | 4.50 |
| HOMO-$E_F$ barrier (eV) | 0.50 | 0.50 |
| Electron affinity EA (eV) | 3.20 | 2.55 |
| Ionization energy IE (eV) | 5.70 | 5.00 |
| HOMO-LUMO gap (eV) | 2.50 | 2.45 |

**Table S5**: Energy levels of PDIF-CN$_2$ and PDI8-CN$_2$ respectively, showing that there are two differences between them, a smaller HOMO-LUMO gap of 0.05 eV for PDI8-CN$_2$ compared to PDIF-CN$_2$, and the shift of all the energetic levels of the former closer to the vacuum level

by 0.65 eV with respect to the later. These measurements are in agreement with the contact resistance estimated in both cases. The gap between the conduction band and the work function of the gold electrode is bigger for PDI8-CN$_2$ than for PDIF-CN$_2$, which explain the higher contact resistance for the former compared to the later.

## 10. THz spectroscopy at room temperature

THz transmission spectra were acquired by means of a Fourier Transform Spectrometer (FTS, Blue Sky Spectroscopy Inc.) optimized for the THz region. This spectrometer is equipped with a Si$_3$N$_4$ blackbody source and a pyroelectric detector (QMC Instruments Ltd.) that collects the transmitted radiation during the interferogram scans. The organic semiconductor multi single-crystal films (PDIF-CN$_2$ and PDI8-CN$_2$) were deposited on 1 cm × 1 cm high-resistivity silicon substrates (thickness: 0.5 mm), since such substrates are transparent and exhibit extremely low dispersion in the THz range. After the FTS chamber was vacuum-pumped to remove the water vapor spectral signatures in the measurements, the transmission spectra were retrieved from the average of 500 interferogram scans (resolution: 2 cm$^{-1}$). Finally, the relative transmission reported in Fig. S25 for each sample was obtained by normalizing the spectra with a reference measurement on a bare silicon substrate. The spectra were acquired on crystal layers of equal thickness (ca. 1 μm).

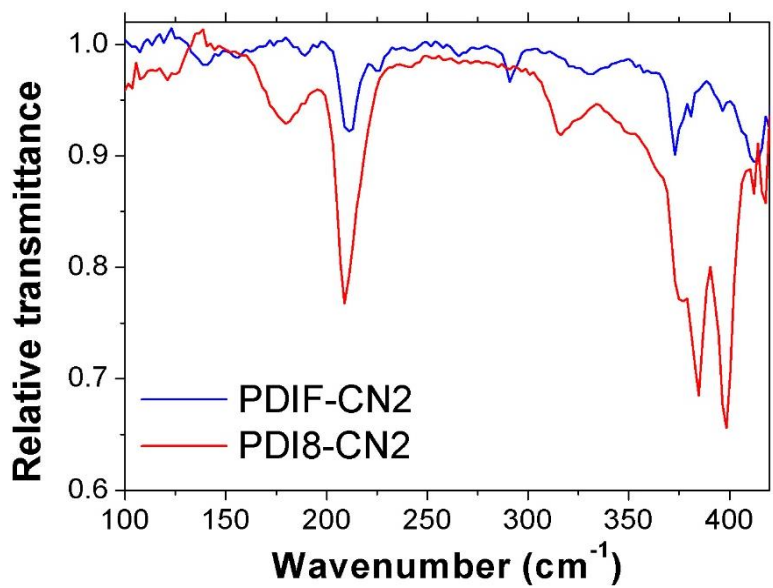

**Figure S37:** Comparative Terahertz spectra of PDIF-CN$_2$ (in blue) and PDI8-CN$_2$ (in red) acquired at T$_{amb}$;